\documentclass[useAMS,usenatbib,usegraphicx]{mn2e}

\title[3-D Magnetic Field Structure in AGN Jets]
{Three-dimensional magnetic field structure of six parsec-scale active galactic nuclei jets}

\author[O'Sullivan \& Gabuzda]{S. P. O'Sullivan$^{1}$ \& D. C. Gabuzda$^{1}$ \\
$^{1}$Physics Department, University College Cork, Cork, Ireland }

\begin{document}

\date{Accepted 2008 November 6. Received 2008 October 30; in original form 2008 May 30}
\pagerange{\pageref{firstpage}--\pageref{lastpage}} \pubyear{2009}
\maketitle
\label{firstpage}
\begin{abstract}
The parsec-scale Faraday rotation measure (RM) distribution of six ``blazars'' (0954+658, 1156+295, 1418+546, 1749+096, 2007+777 \& 2200+420) is investigated using multi-frequency (4.6, 5.1, 7.9, 8.9, 12.9, 15.4, 22 \& 43 GHz) polarization observations taken on 2006 July 2 with the American Very Long Baseline Array (VLBA).
Analysis of the RM provides the direction of the line-of-sight (LoS) magnetic field component, as well as the intrinsic (unrotated) 2-D polarization distribution on the plane of the sky. Our results show that the magnitude of the core RM increases systematically with frequency, and is well described by a power-law, where $|\rm{RM_{core}}$$|$$\propto\nu^{a}$. Our measured values of $a$ vary from 0.9 to 3.8, providing information on the assumed power-law fall-off in the electron density ($n_e$) with distance from the central engine ($r$) for each source.
RM gradients were detected across the jets of 0954+658, 1156+295 and 1418+546, supporting the presence of helical magnetic fields in a sheath or boundary layer surrounding their jets.
We find a bi-modal distribution of the intrinsic jet polarization orientation, with the polarization angles either aligned or orthogonal to the jet direction. The polarization of 2200+420 displays a continuous structure, with the polarization angles remaining aligned with the jet direction even as it bends. This indicates that the magnetic field structure in the synchrotron emitting plasma is dominated by an ordered transverse component. A helical magnetic field geometry can neatly explain both the bi-model distribution of the jet polarization orientation and the ordered polarization structure on these scales.
For 0954+658, 1418+546 and 2200+420, we find that the core RM changes sign with distance from the central engine. We provide an explanation for this by considering a boundary layer of Faraday rotating material threaded by a helical magnetic field, where bends in the relativistic jet or accelerating/decelerating flows give rise to changes in the dominant LoS components of the magnetic field, which in turn gives rise to different signs of the RM.

\end{abstract}
\begin{keywords}
radio continuum: galaxies -- galaxies: jets -- galaxies: magnetic fields
\end{keywords}

\section{Introduction}

Active Galactic Nuclei (AGN) jets emit synchrotron radiation that can be detected at all radio frequencies. This radiation is often highly linearly polarized, which provides important information on the degree of order and orientation of the magnetic field in these jets. The type of AGN we will consider here are known as ``blazars'', which have their jets pointed close to our line of sight (LoS) and often exhibit strong variability in total flux and linear polarization over a broad range of frequencies from $\gamma$-ray to radio. Even though these AGN are intrinsically two-sided, pc-scale observations typically show a one sided ``core-jet'' structure due to Doppler boosting of the radiation from the relativistic jet pointed towards us. The jet moving away from us is usually too faint to be detected.

Recent magneto-hydrodynamic (MHD) simulations have provided an almost complete theoretical picture of how these jets are launched, accelerated and collimated close to the black hole, see \citet{meierjapan} and references therein. These models typically suppose a black hole that collects material in a magnetized, rotating accretion disk. A helical magnetic field is then produced by the differential rotation of the accretion disk or black hole ergosphere, and the magnetocentrifugal forces produced by this field expel material out from the disk along the field lines, while magnetic pinching forces collimate the jet outflow. It is possible that the magnetic field structure is randomised when the flow gets disrupted by shocks after a few hundred Schwarzschild radii, but observational evidence of helical fields on scales larger than this (\citealt{asada2002}; \citealt*{gmc2004}; \citealt{zt2005}; \citealt{mahmudalaska}) suggests that remnants of the earlier magnetic field structure remain, or possibly that a current driven helical kink instability is generated \citep*{nakamurameier2004, nakamurajapan, careyhepro}.

The observed spectra of AGN jets can be optically thick (optical depth, $\tau$, greater than unity), self-absorbed (contribution from both optically thick and optically thin emission regions), or optically thin \citep[e.g.,][]{beckertfalcke2002}. The radio cores of AGN are generally flat-spectrum, self-absorbed regions dominated by emission from around the $\tau=1$ surface and the inner jet. The extended jet emission is generally optically thin, but can have regions of flatter spectra due to particle re-acceleration or interaction with the surrounding medium. Free-free absorption is sometimes detected at low frequencies \citep*[e.g.,][]{gabuzdafreefree}.
In the optically thin regime, the polarization orientation is perpendicular to the magnetic field, with a theoretical maximum degree of polarization of $\simeq 75\%$; in the optically thick regime, the polarization is parallel to the magnetic field and the degree of polarization can be no more than $\simeq 10-15\%$ \citep{pacho1970}.

Multi-frequency VLBI polarization observations allow the effect of Faraday rotation in the immediate vicinity of the AGN to be investigated.
Faraday rotation of the plane of polarization of an electromagnetic wave occurs when it propagates through a region with free electrons and a magnetic field. The effect is manifest as a linear dependence of the observed polarization angle ($\chi_{obs}$) on the wavelength ($\lambda$) squared, described (in SI units) by the formula
\begin{equation}
\chi_{obs}=\chi_0+\frac{e^{3}\lambda^2}{8\pi^2\varepsilon_0 m^2c^3}\int{n}{\bmath{B}\cdot{\bmath{dl}}}\equiv\chi_0+\rm{RM}\lambda^2
\end{equation}
\citep{jackson1975}, where $\chi_0$ is the intrinsic (unrotated) polarization angle and the rotation measure, RM, is proportional to the integral of the Faraday rotating particle density ($n$) and the dot product of the magnetic field ($\bmath{B}$) with the path length ($\bmath{dl}$) along the LoS. Hence, a positive RM tells you that the LoS component of the magnetic field is pointing towards the observer and a negative RM means that the LoS component of the magnetic field is pointing away from the observer.

For a pure electron-positron plasma, one would expect zero net Faraday rotation because of the cancellation of the equal but opposite rotations from the two types of charged particle due to the $e^3$ dependence. The degree of Faraday rotation scales with the inverse square of the effective mass ($\gamma m)^{-2}$, so that the effect is much greater for thermal electrons than for relativistic electrons or protons \citep{jonesodell1977}. The observed Faraday rotation could be from the low energy end of the emitting electron population or from non-radiating thermal electrons possibly entrained in a sheath surrounding the jet \citep{beamsandjets1991}.
However, if the rotating medium is mixed in with the emitting plasma then internal Faraday rotation can occur. Theoretical models of internal Faraday rotation in a spherical or cylindrical region \citep{burn1966, cioffi1980} have shown that, if a rotation of greater than $45\degr$ is observed, then the Faraday rotation must be external. Many VLBI observations show rotations greater than $45\degr$ \citep{zt2003, zt2004, gmc2004, optical2006}, indicating that the main contribution to the rotation must be from gas segregated from the synchrotron emitting region. Furthermore, observations of RM variability \citep{zt2001} and of transverse RM gradients associated with helical magnetic fields \citep[e.g.,][]{asada2002, gmc2004, gabuzdacp2008} indicate that the Faraday rotating gas must be very close to the source, most likely in a boundary layer or sheath surrounding the jet. Therefore, multi-frequency observations enable us to see gas that would otherwise be invisible to us and, combined with the corrected polarization orientation, which corresponds to the magnetic field projected onto the plane of the sky, we are provided with a three dimensional (3-D) view of the magnetic field structure.

A number of previous studies of pc-scale Faraday rotation (\citealt{taylor1998, taylor2000}; \citealt*{reynolds2001}; \citealt{gabuzdachernetskii2003, zt2003, zt2004}) have focused on enhancements in magnitude of the RM in the VLBI core compared to values in the VLBI jet. Some previous results indicated the presence of different RM signs in the core regions of ``blazars'' in different frequency intervals \citep{osullivan2006}. Although RM variability could not be ruled out in this case, the other possibility was that the direction of the dominant LoS magnetic field was changing with distance from the centre of activity. The 5--43 GHz VLBI observations, for which we present results here, were designed to investigate the magnitude and sign of the RM in six AGN jets over a wide range of frequencies/spatial scales.

Section 2 describes the observations and data reduction procedure. Our results are presented in Section 3, followed by their discussion in Section 4. Some conclusions are drawn in Section 5. We assume a cosmology with H$_0 = 71$ km s$^{-1}$ Mpc$^{-1}$, $\Omega_M=0.27$ and $\Omega_{\Lambda}=0.73$. We define the spectral index, $\alpha$, such that $S_{\nu}\propto\nu^{\rm{+}\alpha}$.

\section{Observations and Data Reduction}
We present simultaneous multi-frequency polarization observations obtained on the American Very Long Baseline Array (VLBA) radio interferometer, which provides milliarcsecond (mas) resolution corresponding to the pc-scale structure of these jets, over a 24-hour period on 2006 July 2. Six ``blazars'' (listed in Table \ref{sources}) were observed at eight frequencies  from 4.6 to 43~GHz (Table \ref{evpa}) in a ``snapshot'' mode, with roughly 8--10 scans of each object spread out over the time the source was visible with most or all of the VLBA antennas. This provided uniform $u$--$v$ coverage for all sources. The total data rate for the observation was 128 Mbits s$^{-1}$. The initial calibration was performed with the radio astronomy software package {\sevensize AIPS} using standard techniques. Los Alamos was used as the reference antenna at all stages of the calibration.
The instrumental polarizations (``D-terms'') were derived with the {\sevensize AIPS} task {\sevensize LPCAL} using {\sevensize CLEAN} models for individual features in the compact polarized source 1156+295, which had good parallactic angle coverage over the observational period. After applying the solution to the data, plots of the real vs. imaginary cross-hand polarization data indicated that the ``D-terms'' had been successfully removed.

\begin{table}
 \caption{Observed sources}
 \label{sources}
 \begin{tabular}{@{}lccccccc}
  \hline
  Source (Other Name) & Type & Redshift & Integrated RM \\
                      &      &          & (rad m$^{-2}$) \\
  \hline
  $0954+658$                & BL Lac & $0.368$ & $-15\pm4$\\
  $1156+295$ (4C\,$+29.45$) & Quasar & $0.729$ & $-35\pm1$\\
  $1418+546$                & BL Lac & $0.152$ & $17\pm7$\\
  $1749+096$ (4C\,$+09.57$) & BL Lac & $0.320$ & $95\pm6$\\
  $2007+777$                & BL Lac & $0.342$ & $-20\pm3$\\
  $2200+420$ (BL Lac)       & BL Lac & $0.069$ & $-205\pm6$\\
  \hline
 \end{tabular}
\end{table}

\begin{table}
 \caption{Observed frequencies with corresponding bandwidths and EVPA corrections}
 \label{evpa}
 \begin{tabular}{@{}lcccccc}
  \hline
  Frequency & Bandwidth & $\Delta\chi$ \\
  (GHz)     & (MHz)     & (deg)        \\
  \hline
  $4.608$  & $16$ & $7$   \\
  $5.088$  & $16$ & $11$    \\
  $7.912$  & $32$ & $90$ \\
  $8.879$  & $32$ & $84$  \\
  $12.935$ & $32$ & $62$    \\
  $15.379$ & $32$ & $89$  \\
  $22.232$ & $32$ & $32$  \\
  $43.132$ & $32$ & $88$  \\
  \hline
 \end{tabular}
\end{table}

\begin{figure}
\includegraphics[width=84mm]{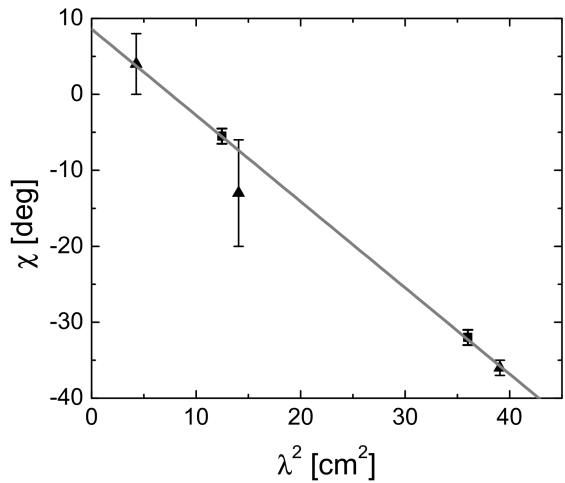}
 \caption{Plot of integrated $\chi$ values, from VLA and UMRAO monitoring of 2200+420, vs. $\lambda^2$. We use 5.0 and 8.5 GHz VLA angles (squares) with 4.8, 8.0 and 14.5 GHz UMRAO angles (triangles) to find an integrated RM of $-198\pm4$ rad m$^{-2}$.}
 \label{integratedRM}
\end{figure}

\begin{table}
 \caption{VLA, UMRAO and VLBA data for the EVPA calibrator 2200+420}
 \label{flux}
 \begin{tabular}{@{}lcccccc}
  \hline
  Telescope & Frequency & Date & $p$ Flux & $\chi$ \\
            & (GHz)     &      & (mJy)    & (deg)   \\
  \hline
  VLA     & $5.0$  & 2006 July 7 & $161$ & $-32.0$   \\
  $ $     & $8.5$  & 2006 July 7 & $142$ & $-5.5$    \\
  $ $     & $22$   & 2006 July 7 & $136$ & $10.5$ \\
  $ $     & $43$   & 2006 July 7 & $106$ & $33.5$  \\
  UMRAO   & $4.8$  & 2006 July 3 & $165$ & $-36.0$    \\
  $ $     & $8.0$  & 2006 June 18 & $143$ & $-13.5$  \\
  $ $     & $14.5$ & 2006 June 26 & $132$ & $4.0$  \\
  \hline
  VLBA    & $4.6$  & 2006 July 2 & $161$ & $-46.5$  \\
  $ $     & $5.1$  & 2006 July 2 & $150$ & $42.5$  \\
  $ $     & $7.9$  & 2006 July 2 & $128$ & $82.0$  \\
  $ $     & $8.9$  & 2006 July 2 & $134$ & $-87.0$  \\
  $ $     & $12.9$ & 2006 July 2 & $126$ & $-59.0$  \\
  $ $     & $15.4$ & 2006 July 2 & $141$ & $-82.0$  \\
  $ $     & $22$   & 2006 July 2 & $133$ & $-21.0$  \\
  $ $     & $43$   & 2006 July 2 & $107$ & $-54.0$  \\
  \hline
 \end{tabular}
\end{table}

The calibration of the electric vector position angles (EVPAs) was determined using integrated polarization measurements from the Very Large Array (VLA) polarization monitoring program\footnote{http://www.aoc.nrao.edu/$\sim$smyers/calibration/} and the 26m radio telescope operated at the University of Michigan Radio Astronomy Observatory (UMRAO). The source 2200+420 was used for our polarization position angle calibration. We used integrated polarization measurements from the VLA on 2006 July 4 at 5, 8.5, 22 and 43 GHz, and from UMRAO at 4.8 (July 3), 8.0 (June 18) and 14.5 GHz (June 26). These observations were compared with our total VLBI polarized flux ($p = \sqrt{Q^2 + U^2}$) and position angle ($\chi = \frac{1}{2}\arctan \frac{U}{Q}$) for 2200+420 on 2006 July 2. Since this calibration scheme relies on the fact that a large fraction of the polarized flux observed by the VLA is also detected with the VLBA (i.e., negligible intermediate scale polarized structure), it is very important for the polarized fluxes to match. Table \ref{flux} shows that, for 2200+420, more than 90\% of the VLA and UMRAO polarization was present on the VLBA scales at all frequencies. Corrections for the EVPAs were then found by aligning the VLA and VLBA polarization position angles ($\Delta\chi=\chi_{Int}-\chi_{obs}$). The polarization monitoring observations were interpolated for our closely spaced frequency pairs in accordance with the $\lambda^2$ law displayed by the integrated EVPAs (see Fig.~\ref{integratedRM}). The integrated RM indicated by these measurements is consistent with the value of $-205\pm6$ rad m$^{-2}$ measured at longer wavelengths by \citet{rudnickjones1983}. These corrections (Table \ref{evpa}), relative to the same reference antenna, appear to be stable across long periods of time, and we estimate that they are accurate to within $\pm3\degr$.

The imaging was performed in {\sevensize AIPS} using ``robust'' weighting, which is a combination of the natural and uniform weighting schemes. This weighting scheme is very useful for VLBA data of AGN jets because good resolution is needed to image the core region but we also want good sensitivity to the extended jet structure. Distributions of the polarized flux ($p$) and polarization angle ($\chi$), as well as accompanying ``noise maps'', were constructed for all sources using the {\sevensize AIPS} task {\sevensize COMB}. The uncertainties written in the output $\chi$ noise maps were calculated in {\sevensize COMB} using the rms noise levels on the input Stokes $Q$ and $U$ maps.

\begin{figure*}
\includegraphics[width=168mm]{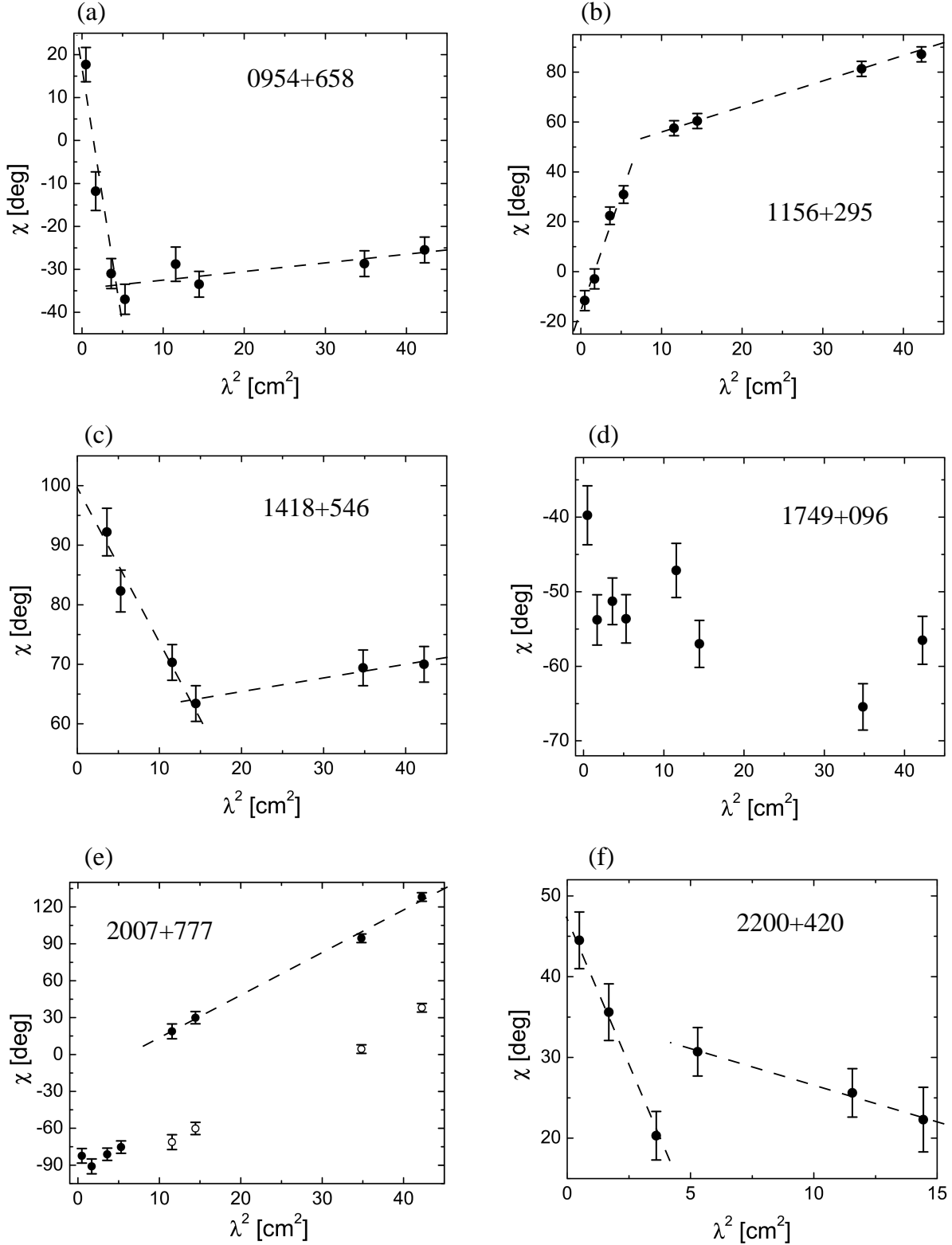}
 \caption{Plots of $\chi_{obs}$ vs. $\lambda^2$ for the whole observed frequency range of the core regions of the full resolution images of 0954+658 (a), 1156+295 (b), 1418+546 (c), 1749+096 (d), 2007+777 (e) and 2200+420 (f). Dotted lines inserted where obvious linear dependences are present. Clear transitions in 0954+658, 1156+295, 1418+546 and 2200+420 outline where physically seperate core RM regions are being probed along the jet. The hollow data points for 2007+777 (e) are rotated by $90\degr$ due to the detection of an optical depth transition in the core between 8.9 and 12.9 GHz. The 4.6 and 5.1 GHz core EVPAs of 2200+420 (f) are not included in the plot in order to more clearly show the core RM transition at the shorter wavelengths; see Fig.~\ref{2200RM}a for the 4.6 and 5.1 GHz angles.}
 \label{RMtransitions}
\end{figure*}

\begin{table*}
 \caption{Spectral indices corresponding to RM ranges}
 \label{spix}
 \begin{tabular}{@{}ccccccc}
  \hline
   &$\alpha_{\rm{core}}$&$\alpha_{\rm{core}}$&$\alpha_{\rm{core}}$&$\alpha_{\rm{jet}}$&$\alpha_{\rm{jet}}$&$\alpha_{\rm{jet}}$ \\
  Source & Low $\nu$ range  & Mid $\nu$ range & High $\nu$ range  & Low $\nu$ range & Mid $\nu$ range & High $\nu$ range \\
  \hline
  $0954+658$ & 0.3 & 0.2 & 0.2 & -0.5 & -0.6 & -0.5 \\
  $1156+295$ & 1.3 & 0.5 & 0.0 & -0.8 & $-$ & $-$ \\
  $1418+546$ & 0.4 & 0.3 & $-$ & -0.2 & -0.8 & $-$ \\
  $1749+096$ & 0.8 & 0.0 & -0.1 & $-$ & $-$ & $-$ \\
  $2007+777$ & 0.6 & 0.3 & 0.2 & -0.4 & -0.6 & -0.6 \\
  $2200+420$ & 0.7 & 0.5 & 0.5 & -0.5 & -0.7 & -0.5 \\
  \hline
 \end{tabular}\\
 \medskip
 \sevensize{Spectral indices for the low, mid and high $\nu$ ranges are derived between the highest and lowest frequencies in the corresponding RM ranges in Table \ref{resultstable}. The errors in the spectral index values are $\pm0.1$.}
\end{table*}

\begin{table*}
 \caption{Summary of core RM results, with all RM values quoted in rad m$^{-2}$.}
 \label{resultstable}
 \begin{tabular}{@{}cccccccccccc}
  \hline
Source &  RM$_{\rm{core}}$     & $m_{\rm{core}}$ & RM$_{\rm{core}}$     & $m_{\rm{core}}$ & RM$_{\rm{core}}$   & $m_{\rm{core}}$ & $a$\\
Name & (Low $\nu$ range)& (4.6 GHz) & (Mid $\nu$ range) & (7.9 GHz) & (High $\nu$ range) & (15.4 GHz)& ($|\rm{RM_{core}}$$|$$\propto\nu^{a}$)\\
\hline
$0954+658$ & $+33\pm14$ & 7.2\% & $-88\pm23$ & 5.1\% & $-1591\pm265$ & 1.2\% & $3.84\pm1.34$\\
$1156+295$ & $+170\pm5$ & 2.6\% & $+618\pm91$ & 2.8\% & $+1667\pm159$ & 1.3\% & $2.22\pm0.05$\\
$1418+546$ & $+83\pm11$ & 4.3\% & $-501\pm48$ & 3.4\% &     $-$     & 1.8\% & $3.32\pm0.60$\\
$1749+096$ & $-$ & 3.3\% & $-$ & 2.9\% & $-$ & 3.3\% & $-$\\
$2007+777$ & $+638\pm39$ & 2.8\% & $-$ & 1.4\% & $+1630\pm201$ & 5.7\% & $0.91\pm0.62$\\
$2200+420$ & $-193\pm29$ & 1.3\% & $+240\pm90$ & 0.9\% & $-1008\pm43$ & 3.6\% & $1.40\pm0.18$\\
  \hline
 \end{tabular}\\
\sevensize{Low $\nu$ range: generally 4.6--8.9 GHz; Mid $\nu$ range: generally 7.9--15.4 GHz; High $\nu$ range: generally 15.4--43 GHz.
Exact ranges (see text and RM maps) depend on where clear RM transitions occur and on the level of polarized flux detected in the jet at higher frequencies. No polarization was detected for 1418+546 at 22 and 43 GHz, hence there is no high $\nu$ range RM for this source. No reliable $\chi_{obs}$ vs. $\lambda^2$ fits were found for 1749+096 across the entire frequency range.
$m_{\rm{core}}$: core degree of polarization at the indicated frequency.
The calculated power-law dependence, $a$, of the core RM with frequency for each source is quoted in the final column.}
\end{table*}

\subsection{Analysis of the Rotation Measure}
In general, the total Faraday rotation for AGN is dominated by two components: (1) The foreground contribution (mainly from our own Galaxy, see e.g., \citet{pushkarev2001} and references therein), and (2) magnetized thermal plasma in the immediate vicinity of the radio source. The foreground contribution is constant over the entire radio source on milli-arcsecond scales, whereas the magneto-ionic Faraday rotating medium near the radio source may be non-uniform on these scales; for example, the magnitude of the RM is usually enhanced in the VLBI core region compared to its value in the VLBI jet \citep[e.g.,][]{zt2003, zt2004}.

Removing the integrated RM is very important for our analysis for two main reasons. (1) The sign of the RM carries information about the direction of the LoS magnetic field giving rise to the Faraday rotation. In this sense, the sign of the RM contains information about the 3-D structure of the magnetic field, which is lost if we are not able to isolate Faraday rotation local to the AGN from the Galactic component. (2) The relationship between the observed RM ($RM_{obs}$) and the RM in the rest frame of the source ($RM_{0}$) is $RM_{obs} = RM_{0}/(1+z)^2$. For AGN with relatively low redshifts, the values of $RM_{obs}$ and $RM_{0}$ will be very close, but for those with higher redshifts, it is essential to remove the foreground RM contribution if we are interested in comparing the intrinsic RMs in different AGN at different redshifts.
Therefore, before making the RM maps with the {\sevensize AIPS} task {\sevensize RM}, we first removed the contribution of the known integrated Faraday rotation at each frequency, so that any remaining RM should be solely due to magnetized thermal plasma in the vicinity of the AGN.

Observations with both long and short frequency spacings are required for reliable Faraday rotation analysis, due to possible $\pm n\pi$ ambiguities in the observed polarization angles. Our ``broadband'' frequency observations (5--43 GHz) enable us to construct RM maps on a range of different scales and Faraday depths, providing a wealth of information on the RM distribution and \emph{sign} in the core region and out along the jet. It's important to note that VLBI resolution is usually not sufficient to completely resolve the true optically thick core, therefore, the VLBI ``core'' consists of emission from the true core and some of the optically thin inner jet. Also, observations at different frequencies are detecting radiation near different $\tau=1$ surfaces, so different frequency intervals probe different physical scales of the inner jet. Hence, matched-resolution images, convolved with the lowest frequency-interval beam, were constructed for different frequency intervals where clear transitions in RM occurred (e.g., Fig.~\ref{RMtransitions}).

For the purposes of determining accurate RM values for individual regions in the VLBI core and jet, we found the mean polarization angle $\chi$ within a $3\times3$~pixel area at the same location in the polarization maps at each frequency. We identify the ``core'' as the furthest upstream self-absorbed jet region, which for some sources, especially at low frequencies, is upstream of the total intensity peak. We arbitrarily choose $3\times3$~pixels ($0.3\times0.3$ mas) as the area used to obtain the EVPA value and an estimate of its rms deviation in the corresponding $\chi$ noise map, which is then added in quadrature with the estimated EVPA calibration uncertainty of $3\degr$. Finally, a reduced $\chi^2$ linear regression was performed to find the RM and its standard error.

\section{Results}

We first outline some general methods and results before considering each individual source.

\subsection{Spectral Index}

Spectral index maps ($S_{\nu} \propto \nu^{\alpha}$) were constructed between pairs of frequencies to obtain detailed spectral information across the entire frequency range. Each map was corrected for the frequency dependence of the optically thick core position using a cross-correlation technique for aligning the optically thin emission from the jet \citep{crokegabuzda2008}. This method is extremely useful for smooth or weak jets that do not have distinct components at each frequency with which to align the images. For sources with very sparse or negligible jet emission, in particular 1156+295 and 1749+096 at high frequencies, we are unable to use this alignment method. In the case of 1156+295, where we have calculated core-shifts at lower frequencies, we predicted the shifts expected at higher frequencies assuming that equipartition holds in these regions \citep[see][]{lobanov1998} and checked that the resulting spectral index maps were consistent. For 1749+096, we simply aligned the images by their peak flux densities and checked that the spectral index maps seemed reasonable.

All sources display a flat or inverted spectrum across all frequencies in the furthest upstream region, consistent with a self-absorbed core.
In general, the jet emission for all sources was optically thin ($\alpha\leq-0.5$), with some flat spectrum regions possibly associated with particle re-acceleration or interaction with the surrounding medium.
Detailed information on the spectra is required in order to identify any regions where possible optically thick-thin transitions occur, requiring the observed polarization angles to be rotated by $90\degr$ for correct Faraday rotation analysis. A detailed analysis of our spectral index maps will be presented in a forthcoming paper. Spectral indices for both core and jet regions corresponding to the RM ranges for each source are summarized in Table \ref{spix}.

\begin{figure*}
\includegraphics[width=168mm]{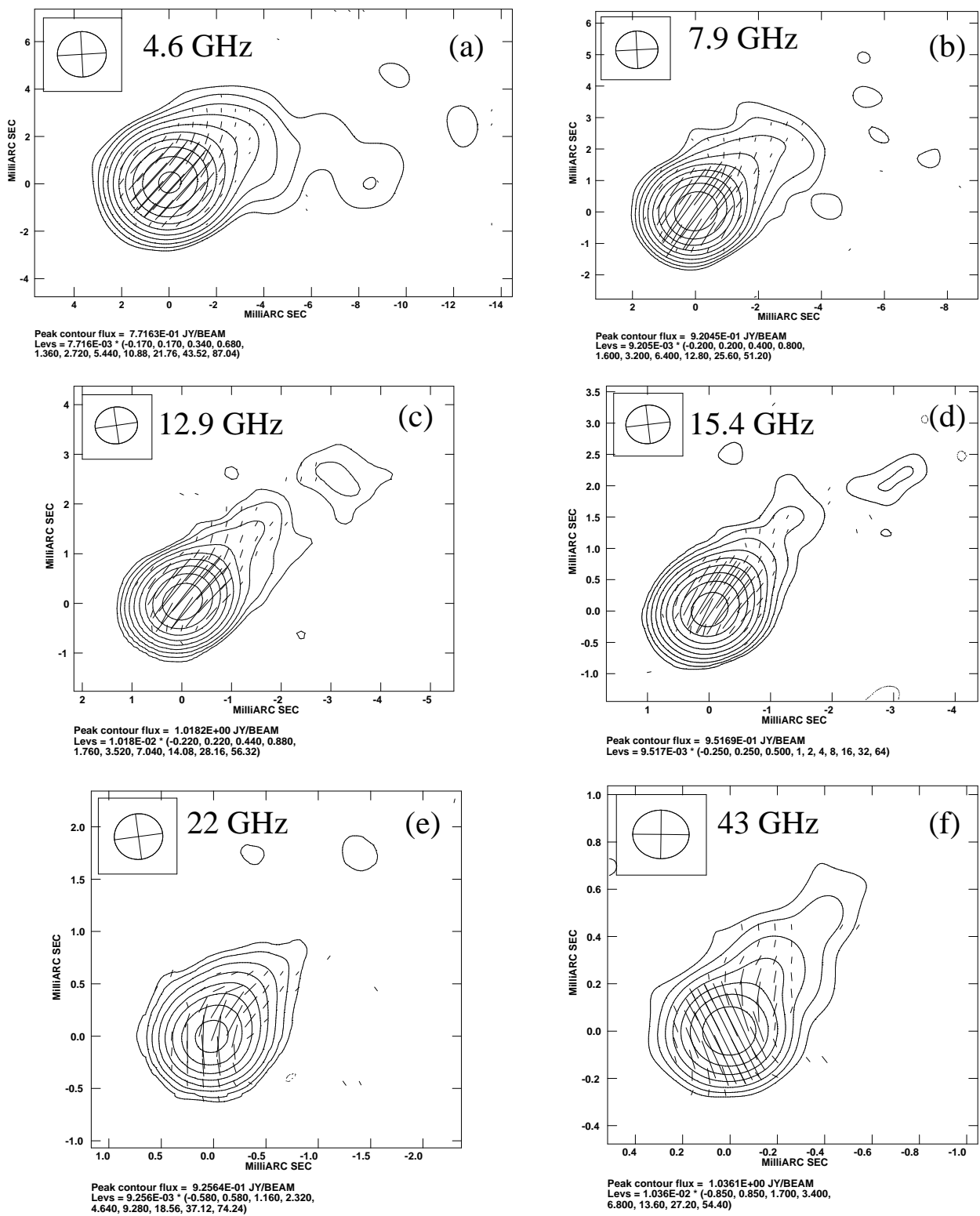}
\caption{Total intensity ($I$) and polarization ($p$) maps for 0954+658: (a) 4.6 GHz; $I$ peak 0.77 Jy bm$^{-1}$; lowest contour 1.31 mJy bm$^{-1}$; $p$ vectors clipped at 1.1 mJy bm$^{-1}$, (b) 7.9 GHz; $I$ peak 0.92 Jy bm$^{-1}$; lowest contour 1.84 mJy bm$^{-1}$; $p$ vectors clipped at 1.1 mJy bm$^{-1}$, (c) 12.9 GHz; $I$ peak 1.02 Jy bm$^{-1}$; lowest contour 2.24 mJy bm$^{-1}$; $p$ vectors clipped at 1.3 mJy bm$^{-1}$, (d) 15.4 GHz; $I$ peak 0.95 Jy bm$^{-1}$; lowest contour 2.38 mJy bm$^{-1}$; $p$ vectors clipped at 1.3 mJy bm$^{-1}$, (e) 22 GHz; $I$ peak 0.93 Jy bm$^{-1}$; bottom contour 5.37 mJy bm$^{-1}$; $p$ vectors clipped at 2.5 mJy bm$^{-1}$, (f) 43 GHz; $I$ peak 1.04 Jy bm$^{-1}$; lowest contour 8.81 mJy bm$^{-1}$; $p$ vectors clipped at 3.5 mJy bm$^{-1}$. The $I$ contours increase by factors of two in all maps. Length of $p$ vectors represent relative polarized intensity.}
 \label{0954_I}
\end{figure*}

\subsection{Rotation Measure}
In the current paradigm, Faraday rotation in the vicinity of the AGN occurs in a boundary layer or sheath of magnetised plasma surrounding the jet (Zavala \& Taylor 2005, Attridge et al. 2005). Our observed RMs are consistent with the rotating medium being external to the emitting region. Linear fits of $\chi$ vs. $\lambda^2$ are found in most cases, with rotations of greater than $45\degr$ observed in several cases. Furthermore, the distribution of the degree of polarization of those with rotations less than $45\degr$ does not fit the corresponding predicted depolarization expected for internal Faraday rotation.

Recent observations \citep{optical2006, osullivan2006, jorstad2007} have suggested that the magnitude of the observed core RMs in AGN increase systematically toward higher frequencies. Since higher frequency observations probe regions closer to the base of the jet, higher magnitude RMs simply imply that the electron density and/or the LoS magnetic field strength is increasing closer to the central engine. The frequency dependence of the core RM can be investigated directly using the relation derived in \citet{jorstad2007}, $|\rm{RM_{core}}$$|$$\propto\nu^{a}$, where $a$ describes an assumed power-law fall-off in the electron density with distance $r$ from the central engine, $n_e\propto r^{-a}$. We assume that the fitted core RM corresponds to the core region defined in the lowest frequency observation used (e.g., $|\rm{RM_{core,\nu_{1}-\nu_{2}}}$$|$$\propto(\nu_2)^{a}, \nu_2<\nu_1$). A summary of our core RM results is presented in Table \ref{resultstable} along with our estimated values of $a$.

\subsection{Intrinsic Polarization Orientation}

After correction for Faraday rotation, we are able to find the intrinsic polarization orientation of the core and jet; hence, we can make firm deductions on the intrinsic magnetic field structure. It's important to note that the intrinsic polarization orientations can vary with frequency-interval simply because different RMs may imply different zero-wavelength ($\chi_0$) angles. To determine the orientation of the furthest upstream polarized component that corresponds to the radio core, we use either the highest frequency-interval $\chi_0$ value or the 43-GHz core polarization angle.

 \begin{figure*}
 \includegraphics[width=168mm]{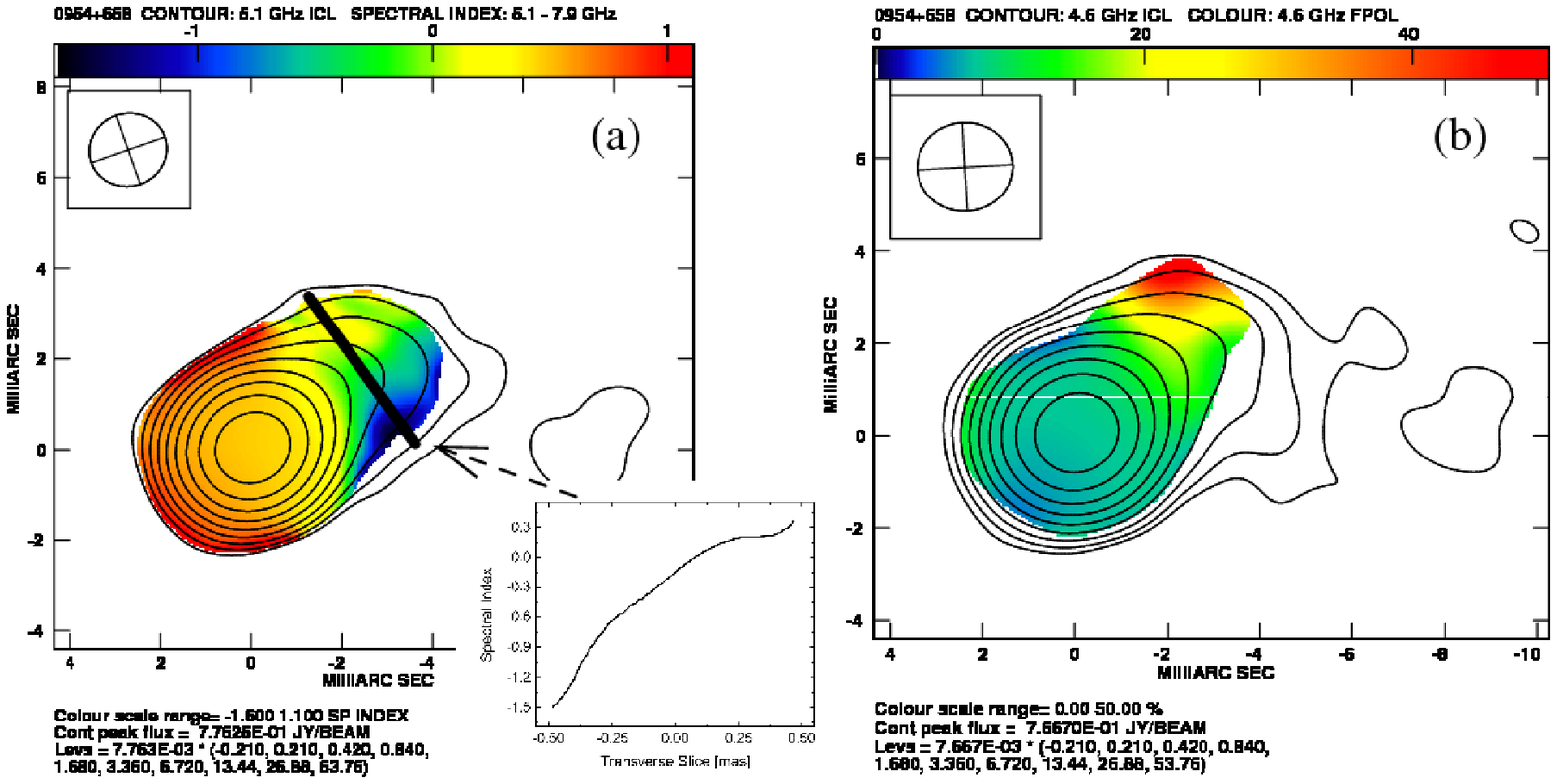}
  \caption{(a) Spectral index map for 0954+658 from 5.1--7.9 GHz ($\alpha$ range: -1.6 to 1.1), overlaid with 5.1 GHz total intensity contours ($I$ peak 0.78 Jy bm$^{-1}$; lowest contour 1.63 mJy bm$^{-1}$). The slice displays the gradient across the jet that may indicate interaction with the surrounding medium where the jet bends. (b) 4.6 GHz total intensity ($I$ peak 0.77 Jy bm$^{-1}$; lowest contour 1.61 mJy bm$^{-1}$) and degree of polarization ($m$), ranging from 0 to 50 percent, for 0954+658 showing how $m$ increases near the region where the jet bends. The $I$ contours increase by factors of two in both maps. Full resolution image in journal version.}
  \label{0954spix+fpol}
 \end{figure*}

\begin{figure*}
\includegraphics[width=162mm]{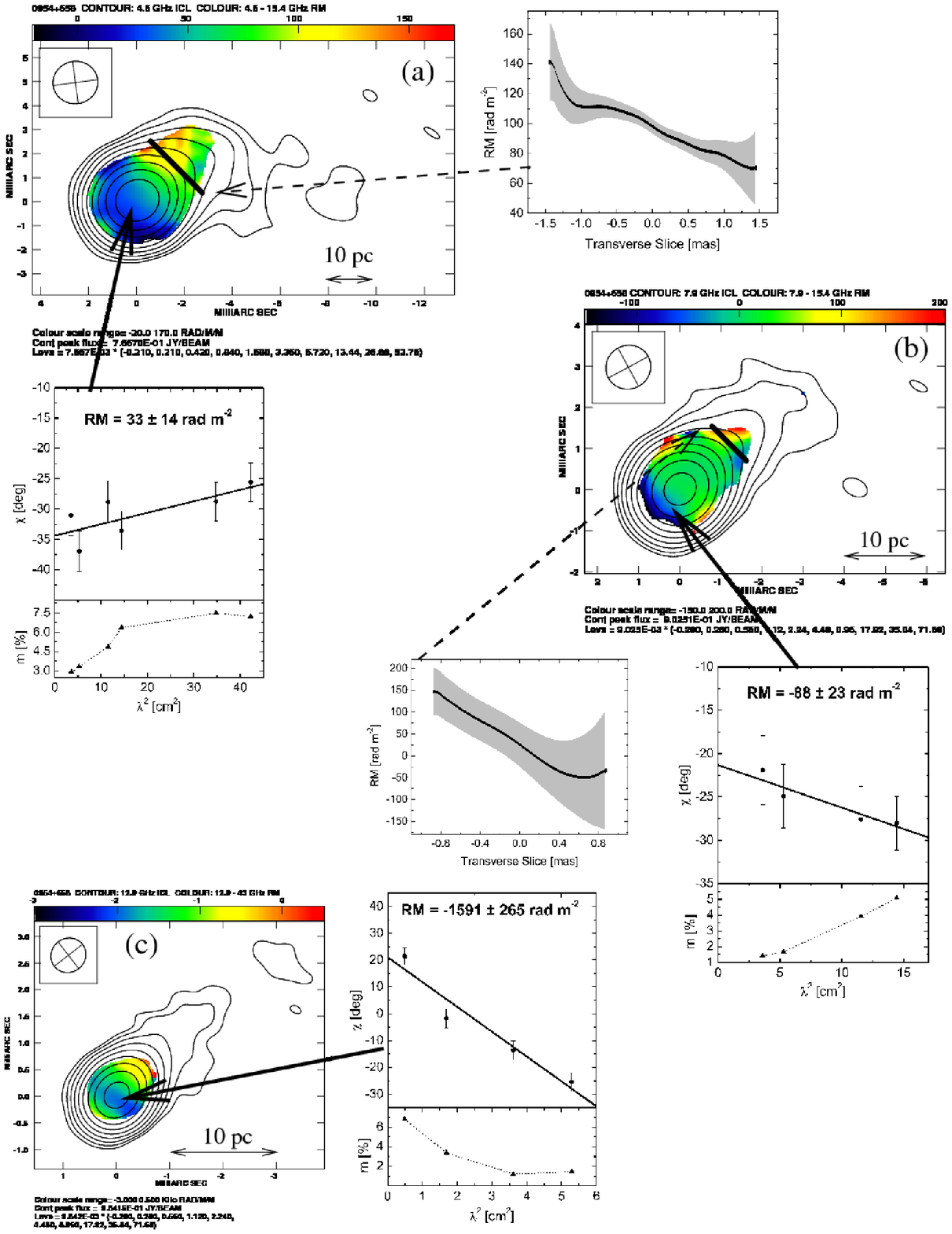}
\caption{Source: 0954+658. (a) 4.6--15.4 GHz RM map (colour scale RM range: -20--170 rad m$^{-2}$) with 4.6 GHz total intensity contours ($I$ peak 0.77 Jy bm$^{-1}$; lowest contour 1.61 mJy bm$^{-1}$). The solid arrow points to the core region where the $\chi$ vs. $\lambda^2$ fits shown in the inset were found; the degree of polarization is plotted underneath for each corresponding polarization angle. The dashed arrow points to the indicated slice taken across the RM map; the plotted black line shows the RM fit at every pixel across the slice with the grey area indicating the error in the fit for each point. (b) 7.9--15.4 GHz RM map (colour scale RM range: -150--200 rad m$^{-2}$) with 7.9 GHz total intensity contours ($I$ peak 0.90 Jy bm$^{-1}$; bottom contour 2.53 mJy bm$^{-1}$). (c) 12.9--43 GHz RM map (colour scale RM range: -3--0.5$\times10^3$ rad m$^{-2}$) with 12.9 GHz total intensity contours ($I$ peak 0.98 Jy bm$^{-1}$; lowest contour 2.76 mJy bm$^{-1}$). All $I$ contours increase by factors of two. Full resolution image in journal version.}
\label{0954RM}
\end{figure*}

\begin{figure}
\includegraphics[width=82mm]{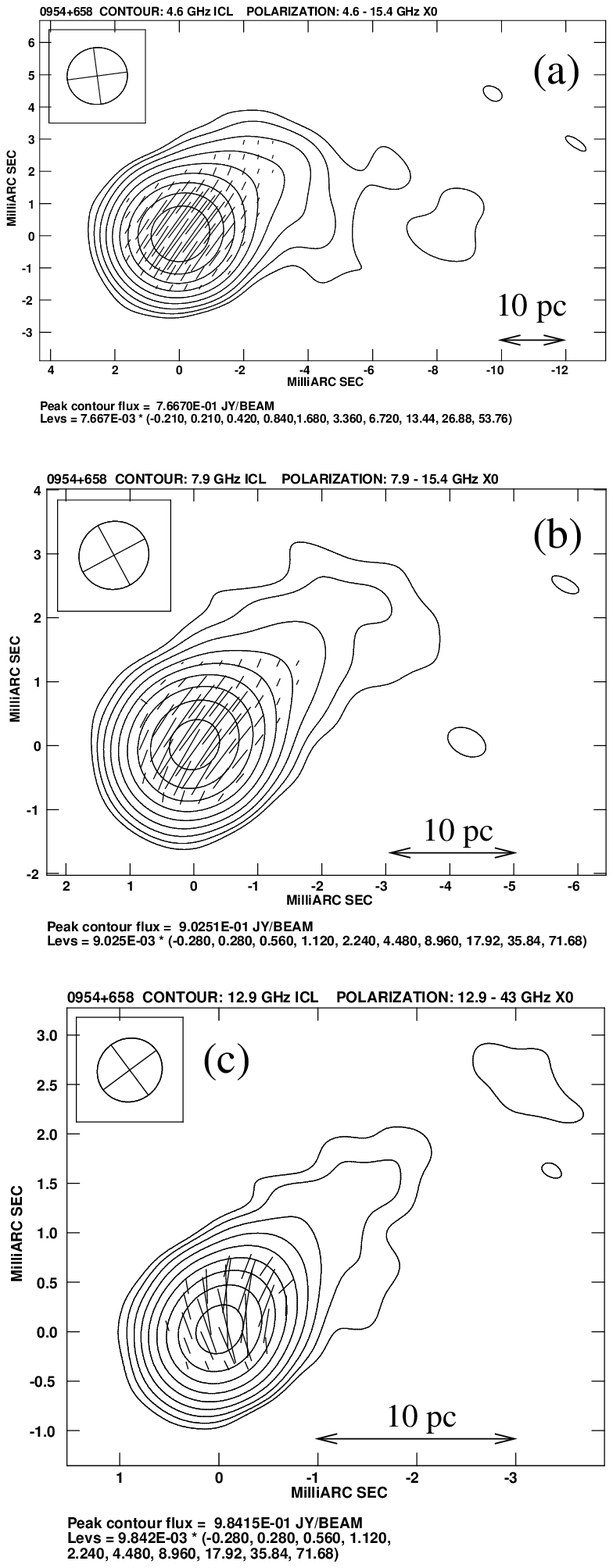}
\caption{Total intensity ($I$) contours with intrinsic polarization orientation ($\chi_0$) for 0954+658 predicted from (a) 4.6--15.4 GHz RM map, (b) 7.9--15.4 GHz RM map and (c) 12.9--43 GHz RM map. Same $I$ contour levels as in Fig.~\ref{0954RM}}
\label{0954X0}
\end{figure}

\subsection{Results for Individual Sources}
We will now consider the results for each source in turn. In each case, maps are presented for six of our eight frequencies; the 5.1 and 8.9~GHz images are omitted due to their similarity to the 4.6 and 7.9~GHz images, respectively.
\subsubsection*{0954+658}
This source is classified as a BL Lac object with a redshift of 0.367. It's a well known Intra-Day Variable (IDV) source \citep[e.g.,][]{0954idv} and has also been detected at $\gamma$-rays by EGRET \citep{mattox2001}. It displays a one-sided kpc-scale jet structure extending to the southwest \citep{cassaro1999}, which seems consistent with a continuation of the observed bend on pc-scales in the 4.6-GHz total intensity map (e.g., Fig.~\ref{0954_I}a). It has an integrated RM of $-15\pm4$ rad m$^{-2}$ \citep{rusk}. We find the VLBI core to be self-absorbed with $\alpha\sim+0.3$ from 4.6--15.4 GHz and $\alpha\sim+0.2$ from 15.4--43 GHz. The jet emission is generally optically thin with $\alpha\sim-0.5$ from 4.6--15.4 GHz and from 15.4--43 GHz (Table \ref{spix}). A notable exception is the detection of a gradient in the 5.1--7.9 GHz spectral index map (Fig.~\ref{0954spix+fpol}a)
where $\alpha$ changes from negative to positive across the jet close to where it bends, possibly indicating interaction with the surrounding medium. Figure \ref{0954_I} displays the total intensity and polarization structure of 0954+658.

The core degree of polarization ($m$) at 4.6 GHz is $\sim7\%$; it then decreases to a minimum of $\sim1.2\%$ at 15.4 GHz before increasing again back up to $\sim7\%$ at 43 GHz. The degree of polarization in the jet is higher and also has a clear gradient across the jet going from $\sim10\%$ at the southern edge to $\sim30\%$ at the northern edge (Fig.~\ref{0954spix+fpol}b). This gradient in $m$ is also seen on scales of 10--20 mas in the same direction \citep{hallahanparis}.

For this source, we were able to produce RM maps on three different scales: from 4.6--15.4, 7.9--15.4 and 12.9--43 GHz. The RM map from 4.6--15.4 GHz (Fig.~\ref{0954RM}a) has a core RM of $33\pm14$ rad m$^{-2}$. There is a clear gradient in RM across the inner jet region before the jet bends; the sign of the RM remains positive across the gradient, varying systematically from $\sim70$ rad m$^{-2}$ to $\sim140$ rad m$^{-2}$ at the northern edge. The magnitude of the RM in the cores of AGN is usually larger than what is found in their jets \citep{zt2003, zt2004}, but in this case the magnitude is lowest in the core. This supports the above evidence from the spectral index map of possible interaction with the surrounding medium where the jet bends, with the higher electron density in this region causing the observed increase in the magnitude of the RM and the increase of the degree of polarization possibly due to ordering of the magnetic field at the site of interaction \citep{aloy2000}. However, the RM and degree of polarization gradients can also be explained in terms of a helical magnetic field structure \citep{lyutikov2005}. This is an example of the difficulty to distinguish between asymmetric RM gradients associated with helcial magnetic fields and/or interaction with the surrounding medium in some cases. See \citet{gomez2008} for a similar scenario in the radio galaxy 3C~120.

Analyzing the RM from 7.9--15.4 GHz (Fig.~\ref{0954RM}b), we find that the core RM changes \emph{sign} and increases in magnitude to $-88\pm23$ rad m$^{-2}$. The inner jet shows some evidence for a continuation of the RM gradient found at lower resolution, however the errors in the fits are very large in this region.

The highest frequency-interval RM map, from 15.4--43 GHz (Fig.~\ref{0954RM}c), only contains information about the core RM, since no appreciable polarization was detected outside the core region at 22 and 43 GHz. The core RM of $-1591\pm265$ rad m$^{-2}$ has the same sign as the 7.9--15.4 GHz value but its magnitude has increased dramatically. Fitting the three core RM values vs. frequency using a least-squares method, we find a large value of $a=3.84\pm1.34$ for the fall-off in electron density from the central engine ($n_e\propto r^{-a}$).

After correcting for Faraday rotation, the intrinsic polarization orientation derived from 4.6--15.4 and 7.9--15.4 GHz, in both the core and jet, is approximately aligned with the jet direction (Fig.~\ref{0954X0}a, b). However, at the highest frequency interval, 15.4--43 GHz, the furthest upstream polarized component is approximately transverse to the jet direction (Fig.~\ref{0954X0}c). The 43-GHz polarization map clearly resolves this component (Fig.~\ref{0954_I}f).

\subsubsection*{1156+295}
\begin{figure*}
\includegraphics[width=162mm]{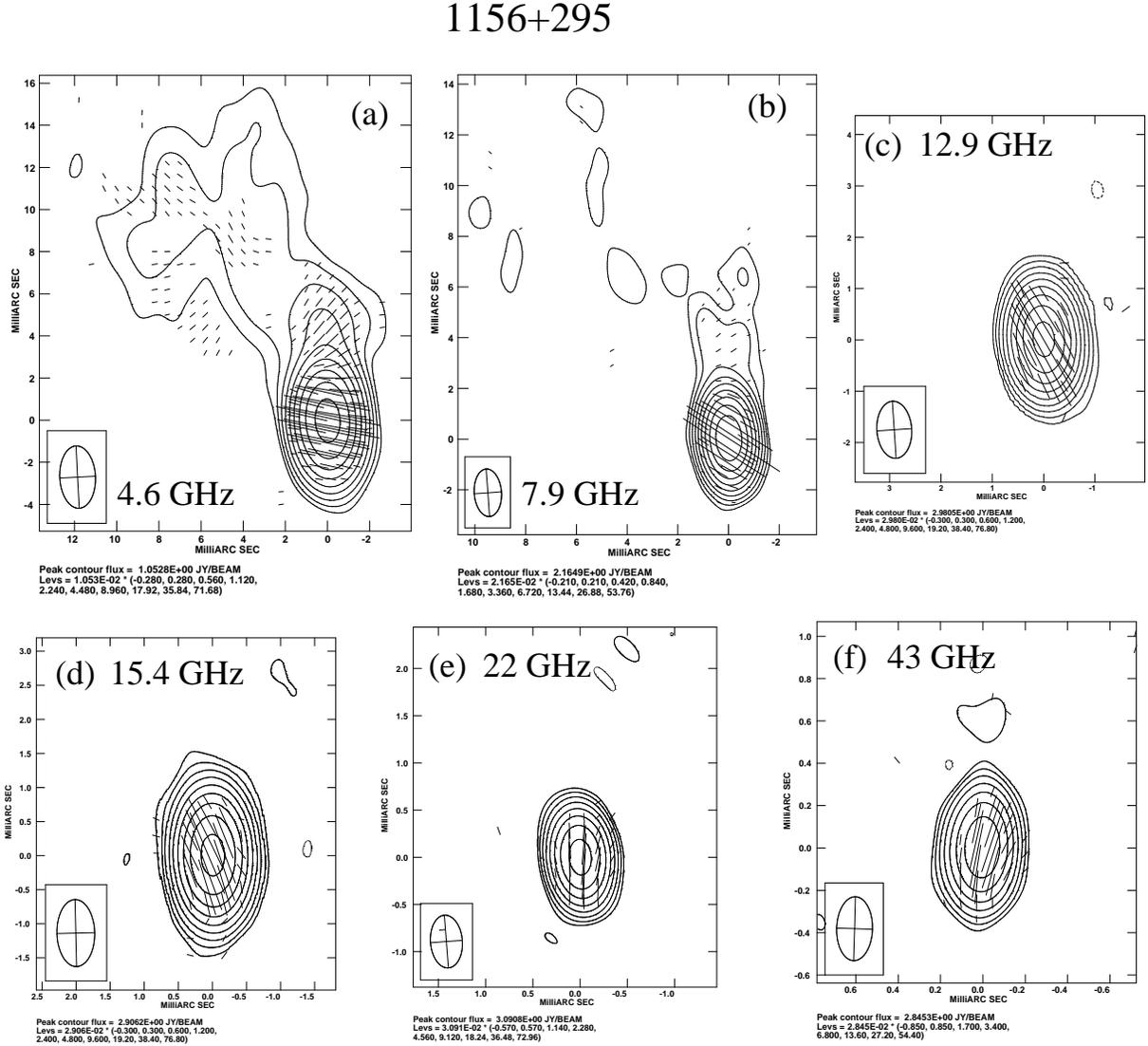}
 \caption{Total intensity ($I$) and polarization ($p$) maps for 1156+295: (a) 4.6 GHz; $I$ peak 1.05 Jy bm$^{-1}$; lowest contour 2.95 mJy bm$^{-1}$; $p$ vectors clipped at 1.1 mJy bm$^{-1}$, (b) 7.9 GHz; $I$ peak 2.16 Jy bm$^{-1}$; lowest contour 4.55 mJy bm$^{-1}$; $p$ vectors clipped at 2.5 mJy bm$^{-1}$, (c) 12.9 GHz; $I$ peak 2.98 Jy bm$^{-1}$; lowest contour 8.94 mJy bm$^{-1}$; $p$ vectors clipped at 2.8 mJy bm$^{-1}$, (d) 15.4 GHz; $I$ peak 2.91 Jy bm$^{-1}$; lowest contour 8.72 mJy bm$^{-1}$; $p$ vectors clipped at 2.8 mJy bm$^{-1}$, (e) 22 GHz; $I$ peak 3.09 Jy bm$^{-1}$; bottom contour 17.62 mJy bm$^{-1}$; $p$ vectors clipped at 3.5 mJy bm$^{-1}$, (f) 43 GHz; $I$ peak 2.85 Jy bm$^{-1}$; lowest contour 24.18 mJy bm$^{-1}$; $p$ vectors clipped at 5.5 mJy bm$^{-1}$. The $I$ contours increase by factors of two in all maps. Length of $p$ vectors represent relative polarized intensity.}
 \label{1156_I}
\end{figure*}

Also known as 4C\,+29.45, this object has been classified as both a High Polarization Radio Quasar (HPRQ) and an Optical Violent Variable (OVV) \citep[see][]{hong2004}. It is the most distant object in our sample of six sources, with a redshift of 0.729. It has also displayed intra-day variability \citep{1156idv2007} and has a $\gamma$-ray detection from EGRET \citep{mattox2001}. On kpc-scales it has a two-sided north-south halo \citep{antonucci1985}; on pc-scales the jet initially propagates directly north and then after a few mas changes to a northeasterly direction, as can be seen in our 4.6-GHz image (Fig.~\ref{1156_I}a). We subtracted an integrated RM of $-35\pm1$ rad m$^{-2}$ \citep{orenwolfe1995} before making the RM maps. From our data, the spectral index in the core is inverted at low frequencies with $\alpha\sim+1.3$ from 4.6--8.9 GHz, indicating strong synchrotron self-absorption. The inner jet is optically thin with $\alpha\sim-0.8$. The core spectrum flattens at higher frequencies with $\alpha\sim0.0$ between 12.9 and 43 GHz (Table \ref{spix}). Figure \ref{1156_I} displays the total intensity and polarization structure of 1156+295.

\begin{figure*}
\includegraphics[width=162mm]{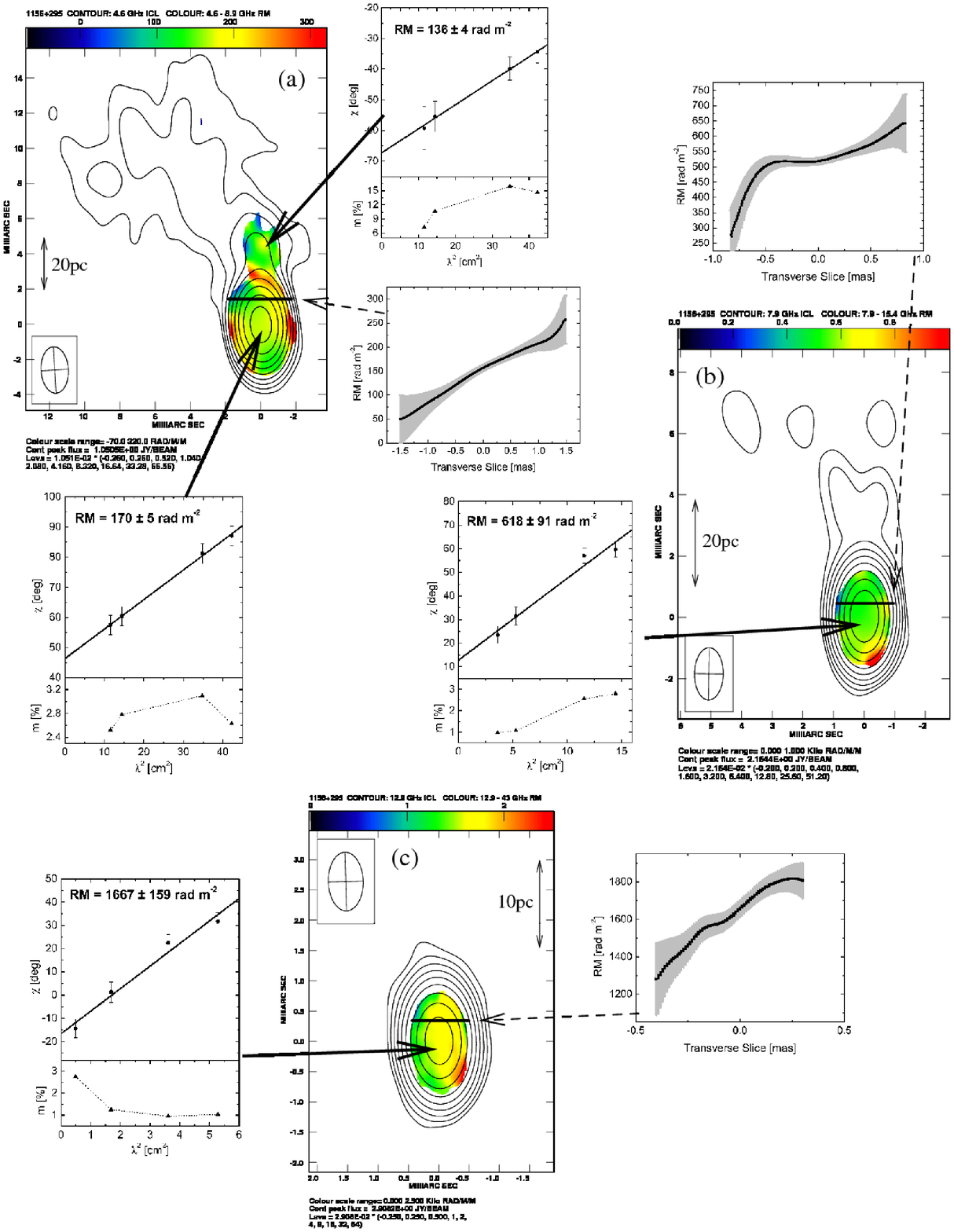}
 \caption{Source: 1156+295. (a) 4.6--8.9 GHz RM map (colour scale RM range: -70--320 rad m$^{-2}$) with 4.6 GHz total intensity contours ($I$ peak 1.05 Jy bm$^{-1}$; lowest contour 2.73 mJy bm$^{-1}$). The solid arrow points to the core region where the $\chi$ vs. $\lambda^2$ fits shown in the inset were found; the degree of polarization is plotted underneath for each corresponding polarization angle. The dashed arrow points to the indicated slice taken across the RM map; the plotted black line shows the RM fit at every pixel across the slice with the grey area indicating the error in the fit for each point. (b) 7.9--15.4 GHz RM map (colour scale RM range: 0--1$\times10^3$ rad m$^{-2}$) with 7.9 GHz total intensity contours ($I$ peak 2.16 Jy bm$^{-1}$; lowest contour 4.33 mJy bm$^{-1}$). (c) 12.9--43 GHz RM map (colour scale RM range: 0--2.5$\times10^3$ rad m$^{-2}$) with 12.9 GHz total intensity contours ($I$ peak 2.91 Jy bm$^{-1}$; lowest contour 7.27 mJy bm$^{-1}$). All $I$ contours increase by factors of two. Full resolution image in journal version.}
 \label{1156RM}
\end{figure*}

$\chi$ vs. $\lambda^2$ fits for the core region were found in three separate frequency ranges: 4.6--8.9, 7.9--15.4 and 12.9--43 GHz (Fig.~\ref{1156RM}). The core RM increases systematically with frequency from $170\pm5$ rad m$^{-2}$ at 4.6--8.9 GHz, to $618\pm91$ rad m$^{-2}$ at 7.9--15.4 GHz and finally to $1667\pm159$ rad m$^{-2}$ at 12.9--43 GHz. This implies a fall-off in the electron density from the central engine of $n_e\propto r^{-2.22\pm0.05}$. There is no detection of a change in sign of the core RM in different frequency intervals. The only jet RM, derived at the lowest frequency interval, of $136\pm4$ rad m$^{-2}$, is smaller in magnitude than the RM in the core. There is evidence for a RM gradient across the inner jet in all three RM maps, as shown through the plots of the sliced regions in Fig.~\ref{1156RM}, confirming the results of \citet{gabuzdacp2008}. The gradient increases from $\sim50$ rad m$^{-2}$ at the eastern edge to $\sim250$ rad m$^{-2}$ at the western edge of the jet in the 4.6--8.9 GHz RM map, from $\sim250$ rad m$^{-2}$ to $\sim650$ rad m$^{-2}$ in the 7.9--15.4 GHz map and from $\sim1300$ rad m$^{-2}$ to $\sim1800$ rad m$^{-2}$ in the 12.9--43 GHz map.

The core degree of polarization decreases from $\sim2.8\%$ at 43 GHz to a minimum of $\sim1\%$ at 15.4 GHz and back up to $\sim3\%$ at 5.1 GHz with a slight decrease to $\sim2.6\%$ at 4.6 GHz. The degree of polarization is $\sim15\%$ in the inner jet at 5.1 and 4.6 GHz.

\begin{figure*}
\includegraphics[width=170mm]{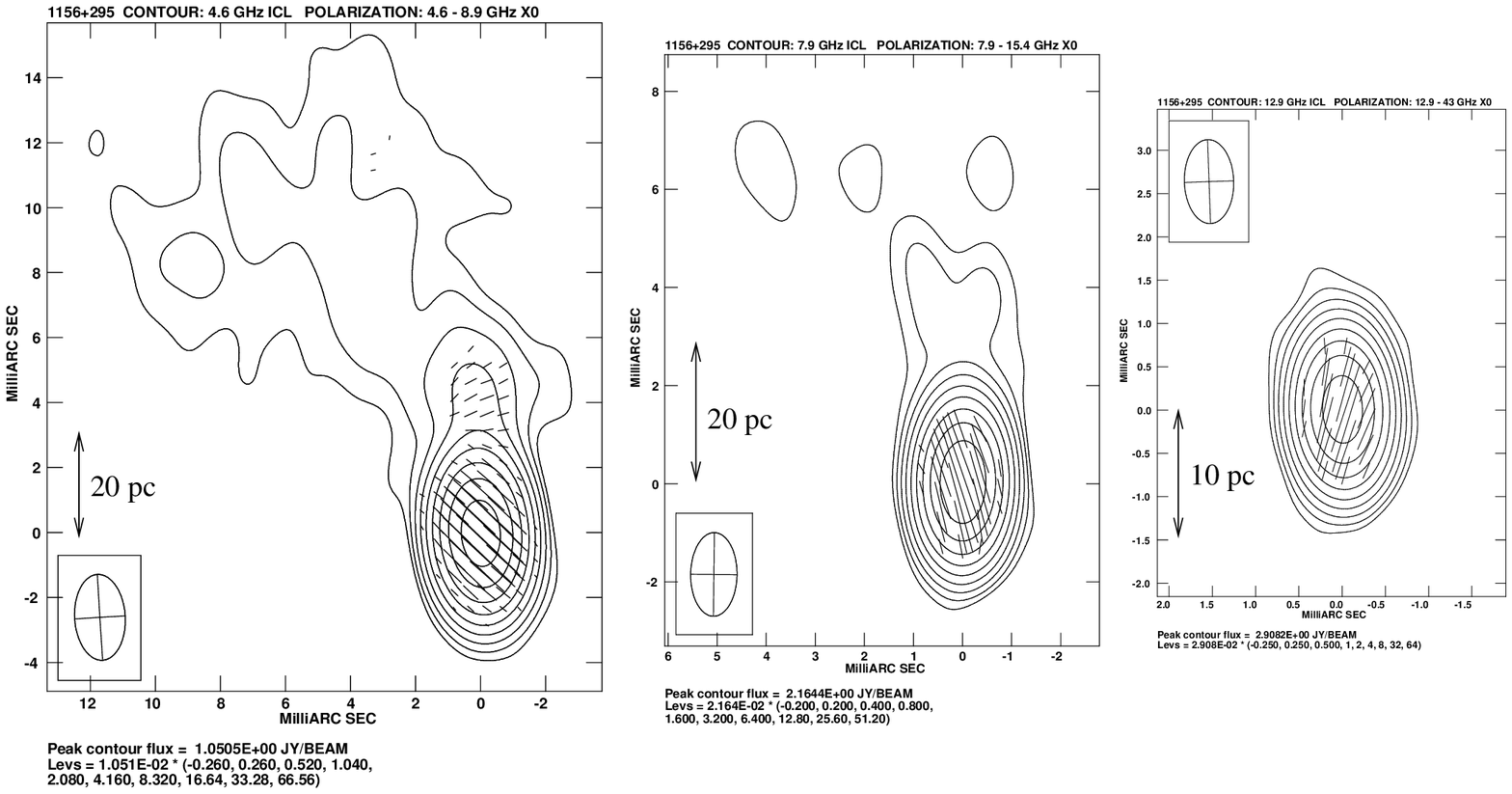}
 \caption{Total intensity ($I$) contours with intrinsic polarization orientation ($\chi_0$) for 1156+295 predicted from (a) 4.6--8.9 GHz RM map, (b) 7.9--15.4 GHz RM map and (c) 12.9--43 GHz RM map. Same $I$ contour levels as in Fig.~\ref{1156RM}}
 \label{1156X0}
\end{figure*}

The outer jet polarization in the 4.6-GHz map displays a spine-sheath type structure with polarization aligned with the jet direction in the central region and transverse to it along the edges of the jet (Fig.~\ref{1156_I}a). In the inner jet region, the Faraday corrected polarization vectors are transverse to the jet direction (Fig.~\ref{1156X0}a) but this is likely to correspond to the sheath polarization structure. The intrinsic core EVPA derived from the highest frequency interval (12.9--43 GHz) is $\sim-20\degr$ which appears to be slightly offset from the jet direction (Fig.~\ref{1156X0}c). However, a jet direction of $\sim-20\degr$ at higher frequencies would not be inconsistent with the helical jet trajectory proposed by \citet{hong2004}.

\begin{figure*}
\includegraphics[width=168mm]{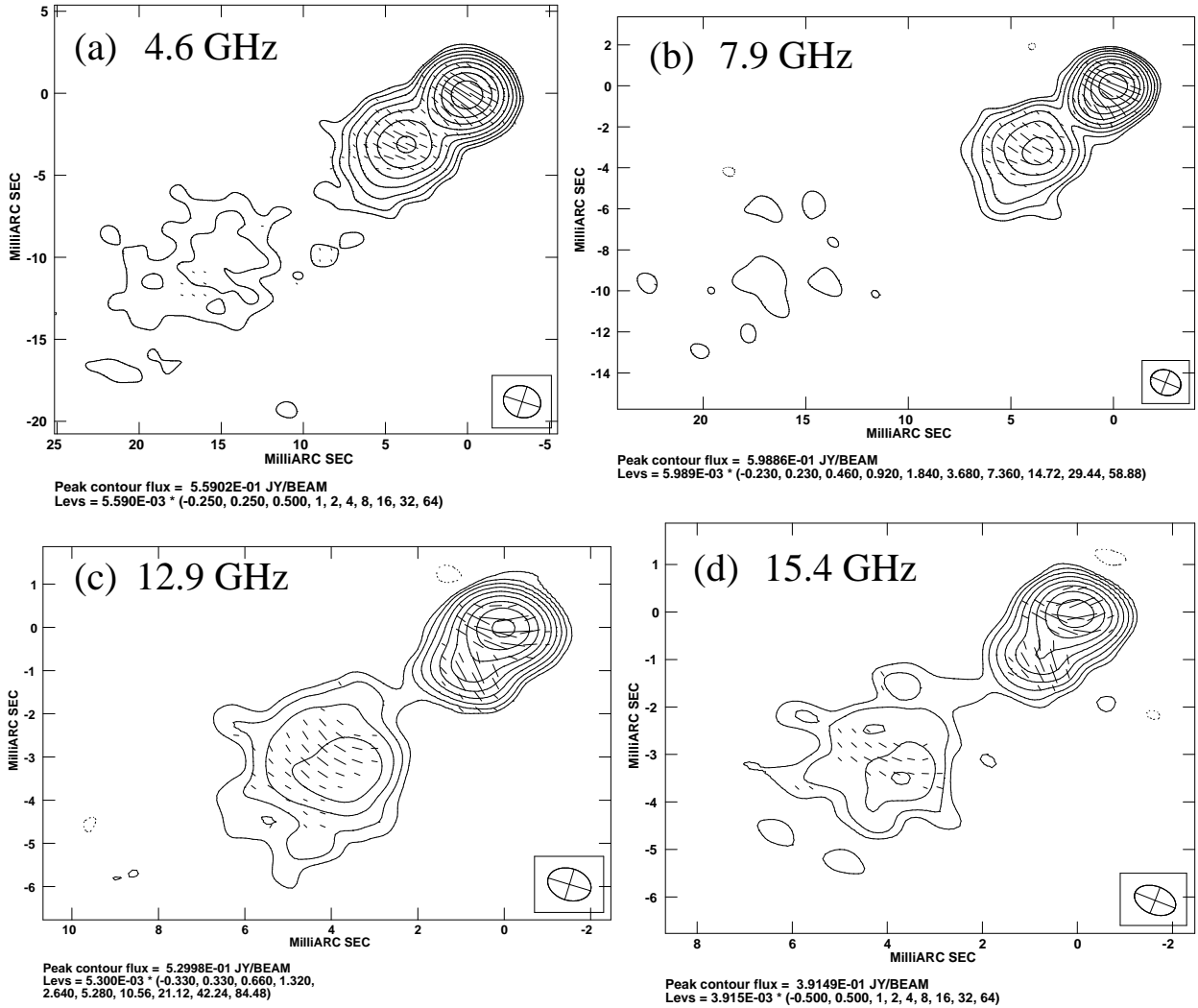}
 \caption{Total intensity ($I$) and polarization ($p$) maps for 1418+546: (a) 4.6 GHz; $I$ peak 0.56 Jy bm$^{-1}$; lowest contour 1.40 mJy bm$^{-1}$; $p$ vectors clipped at 1.0 mJy bm$^{-1}$, (b) 7.9 GHz; $I$ peak 0.59 Jy bm$^{-1}$; lowest contour 1.38 mJy bm$^{-1}$; $p$ vectors clipped at 1.0 mJy bm$^{-1}$, (c) 12.9 GHz; $I$ peak 0.53 Jy bm$^{-1}$; lowest contour 1.75 mJy bm$^{-1}$; $p$ vectors clipped at 1.5 mJy bm$^{-1}$, (d) 15.4 GHz; $I$ peak 0.39 Jy bm$^{-1}$; lowest contour 1.96 mJy bm$^{-1}$; $p$ vectors clipped at 1.5 mJy bm$^{-1}$. The $I$ contours by factors of two in all maps. Length of $p$ vectors represent relative polarized intensity.}
 \label{1418_I}
\end{figure*}

\begin{figure*}
\includegraphics[width=162mm]{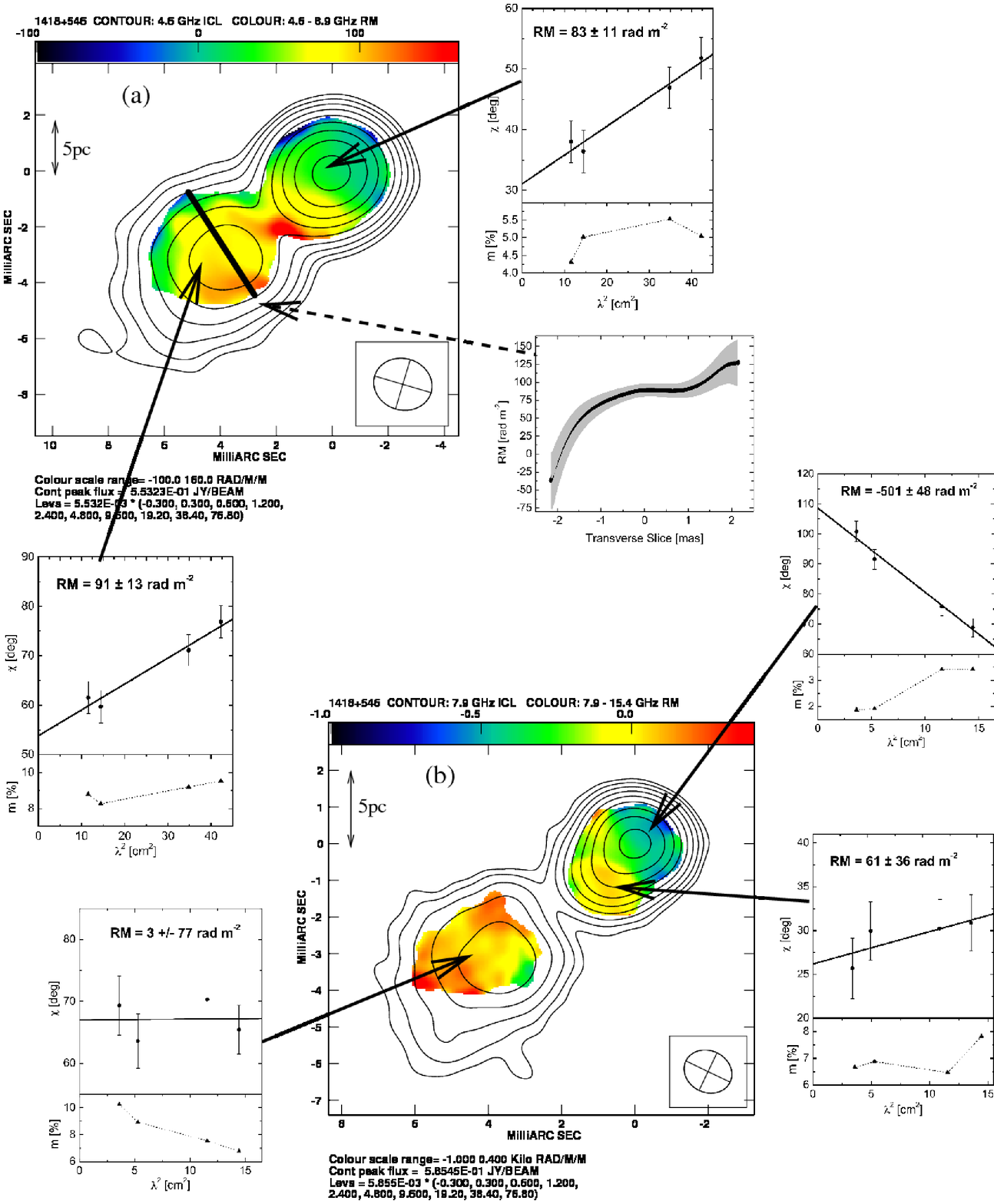}
 \caption{Source: 1418+546. (a) 4.6--8.9 GHz RM map (colour scale RM range: -100--160 rad m$^{-2}$) with 4.6 GHz total intensity contours ($I$ peak 0.55 Jy bm$^{-1}$; lowest contour 1.66 mJy bm$^{-1}$). The solid arrow points to the core region where the $\chi$ vs. $\lambda^2$ fits shown in the inset were found; the degree of polarization is plotted underneath for each corresponding polarization angle. The dashed arrow points to the indicated slice taken across the RM map; the plotted black line shows the RM fit at every pixel across the slice with the grey area indicating the error in the fit for each point. (b) 7.9--15.4 GHz RM map (colour scale RM range: -1--0.4$\times10^3$ rad m$^{-2}$) with 7.9 GHz total intensity contours ($I$ peak 0.59 Jy bm$^{-1}$; lowest contour 1.76 mJy bm$^{-1}$). All $I$ contours increase by factors of two. Full resolution image in journal version.}
 \label{1418RM}
\end{figure*}

\begin{figure*}
\includegraphics[width=168mm]{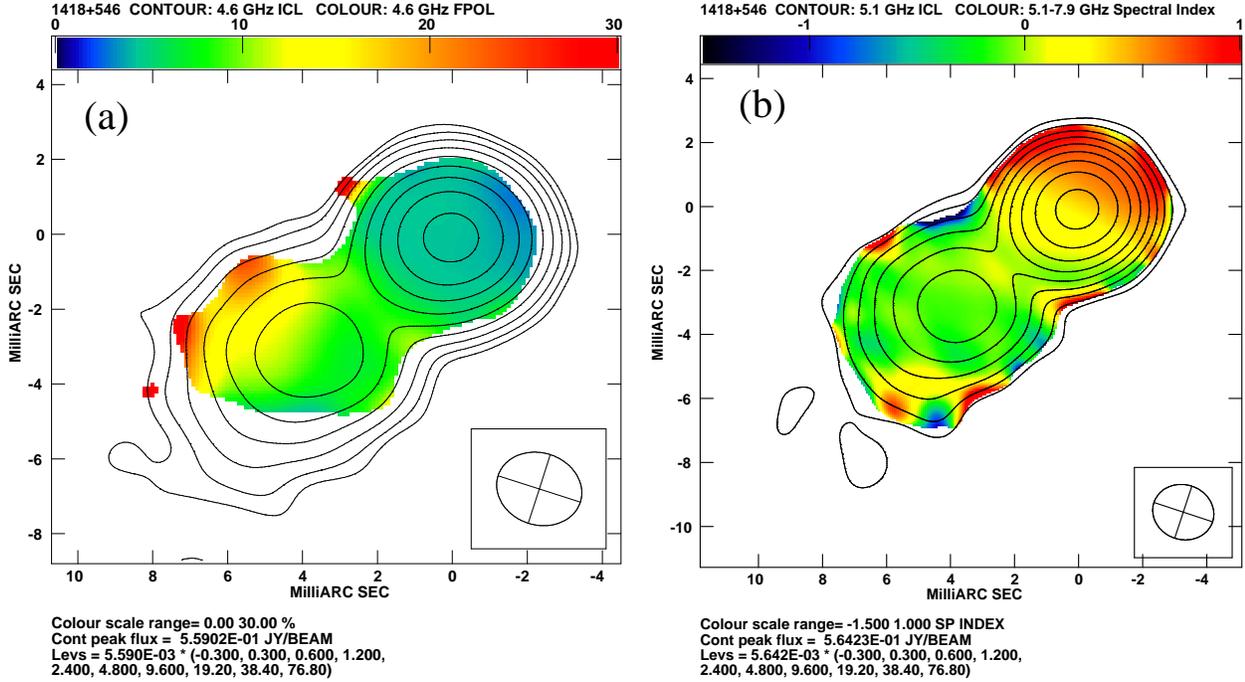}
 \caption{(a) Map of the degree of polarization ($m$), ranging from 0 to 30 percent, at 4.6 GHz for 1418+546 showing a gradient across the jet ($I$ peak 0.56 Jy bm$^{-1}$; lowest contour 1.68 mJy bm$^{-1}$). (b) Map showing spectral index ($\alpha$ range: -1.5 to 1) between 5.1 and 7.9 GHz, with 5.1 GHz contours ($I$ peak 0.56 Jy bm$^{-1}$; lowest contour 1.69 mJy bm$^{-1}$).}
 \label{1418fpol+spix}
\end{figure*}

\begin{figure*}
\includegraphics[width=170mm]{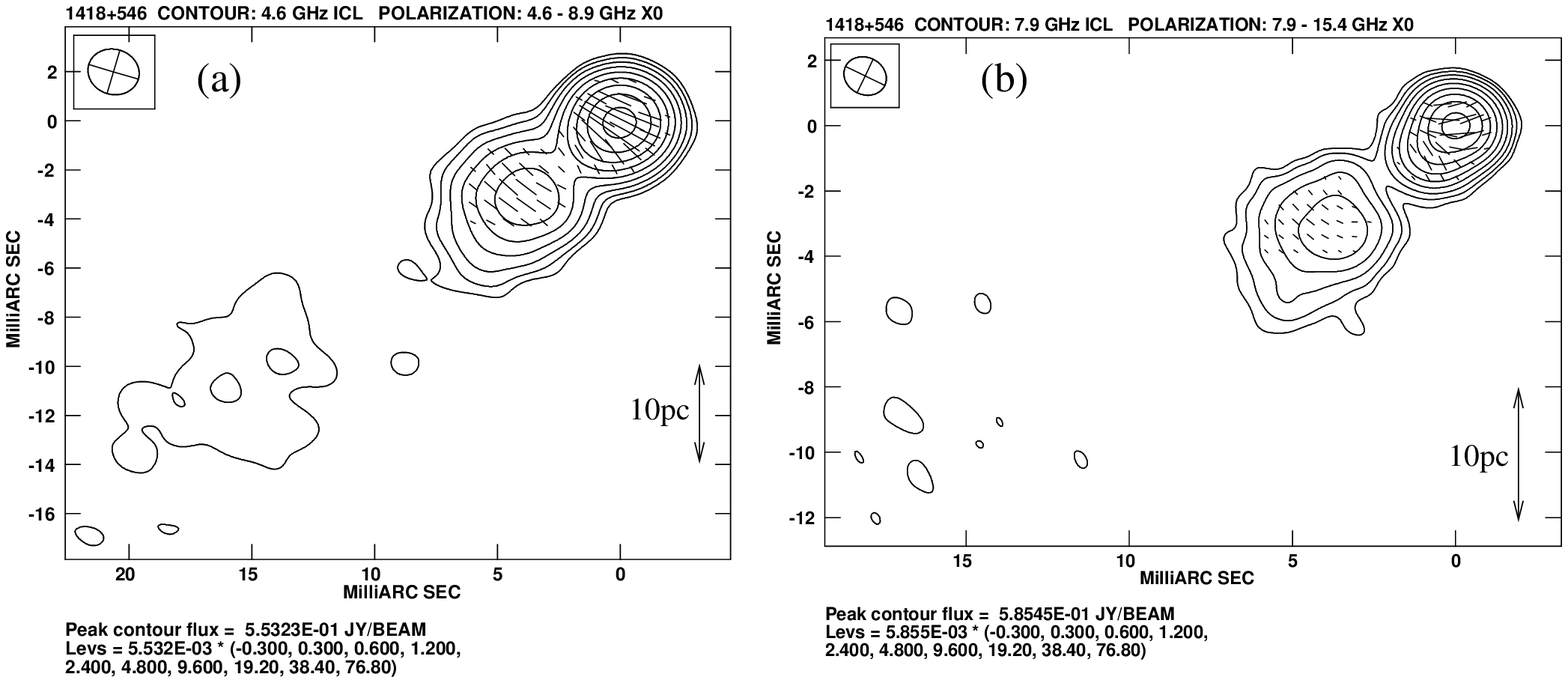}
 \caption{Total intensity ($I$) contours with intrinsic polarization orientation ($\chi_0$) for 1418+546 predicted from (a) 4.6--8.9 GHz RM map and (b) 7.9--15.4 GHz RM map. Same $I$ contour levels as in Fig.~\ref{1418RM}}
 \label{1418X0}
\end{figure*}

\subsubsection*{1418+546}
This BL Lac object, at a redshift of 0.152, has a one-sided kpc-scale halo with a component west of the compact core \citep{cassaro1999} and is misaligned with the VLBI pc-scale jet, which propagates in a south-easterly direction (Fig.~\ref{1418_I}). \citet{rudnickjones1983} found an integrated RM of $17\pm7$ rad m$^{-2}$ for this source. We find a core spectral index of $\alpha\sim+0.4$ consistent with self-absorption and a jet spectral index of $\alpha\sim-0.2$ from 4.6--8.9 GHz. The core region is resolved into two components at higher frequencies and the spectral index map from 7.9--15.4 GHz gives $\alpha\sim+0.3$ for the furthest upstream component, with $\alpha\sim-0.6$ and $\alpha\sim-0.8$ for the inner and outer jet regions, respectively (Table \ref{spix}). Figure \ref{1418_I} displays the total intensity and polarization structure of 1418+546 at four of the six frequencies at which polarization was detected.

We could only obtain RM maps for two frequency intervals for this source (4.6--8.9 and 7.9--15.4 GHz) due to a failure to detect polarization at 22 and 43 GHz. The core RM from 4.6--8.9 GHz is $83\pm11$ rad m$^{-2}$; a RM of $91\pm13$ rad m$^{-2}$ was found in the jet from the same frequency range (Fig.~\ref{1418RM}a). There is also a gradient in RM across the jet from $\sim-30$ rad m$^{-2}$ at the north eastern edge to $\sim120$ rad m$^{-2}$ at the south western edge (Fig.~\ref{1418RM}a). In the higher frequency interval, from 7.9--15.4 GHz, the RM in the core changes \emph{sign} and increases in magnitude to $-501\pm48$ rad m$^{-2}$, giving an estimate of $a\sim3.3$. The RM in the inner jet has a value of $61\pm36$ rad m$^{-2}$ and a zero RM (within the errors) is found in the outer jet region (Fig.~\ref{1418RM}b); there seems to be an apparent RM gradient in this map, however the errors in the higher frequency angles are large and the gradient is not consistently present along the outer jet, unlike the lower frequency RM gradient in approximately this same location. Hence, we don't consider this to represent a real RM gradient, only patchiness in the RM distribution.

The degree of polarization in the core increases steadily from $\sim1.8\%$ at 15.4 GHz to $\sim4.3\%$ at 4.6 GHz. There is a gradient in $m$ across the jet ranging from $\sim5\%$ at the south western edge to $\sim25\%$ at the north eastern edge (Fig.~\ref{1418fpol+spix}a). Since the magnitude of the RM gradient increases in the opposite direction across the jet, and the spectral index map shows no transverse structure in this region (Fig.~\ref{1418fpol+spix}b), interaction with the surrounding medium is unlikely to be a dominant effect in this case. This provides strong support for the direct association of the transverse RM gradient with the underlying magnetic field geometry.

The dominant Faraday corrected intrinsic jet EVPAs are transverse to the jet direction (Fig.~\ref{1418X0}a, b). The polarized emission in the outer jet is clearly offset to the north-eastern edge of the jet from the total intensity emission (e.g., Fig.~\ref{1418_I}). This is similar to the polarization structure in the 5-GHz observations of \cite{pushkarev2004}, where the jet polarization is also offset to the northern edge of the jet. (However, note that, at another epoch, they observed the same jet region to have polarization offset to the southern edge of the jet at both 5 and 8.4 GHz.)
From 4.6--8.9 GHz, the intrinsic EVPAs in the core are also approximately transverse to the jet direction (Fig.~\ref{1418X0}a). However, the higher frequency interval (7.9--15.4 GHz) core $\chi_0$ shows polarization in the furthest upstream component approximately aligned with the jet direction (Fig.~\ref{1418X0}b).

\begin{figure*}
\includegraphics[width=168mm]{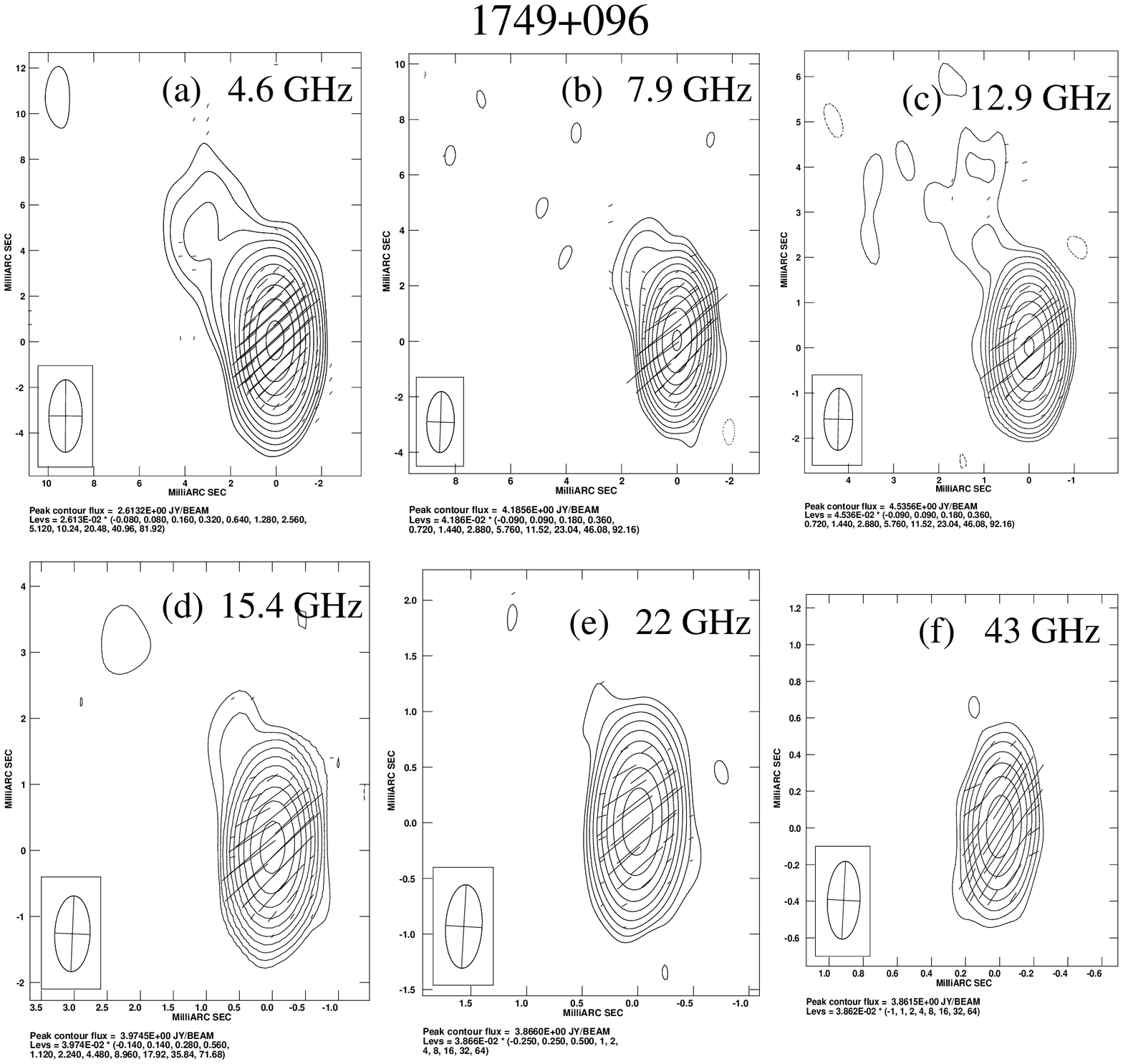}
 \caption{Total intensity ($I$) and polarization ($p$) maps for 1749+096: (a) 4.6 GHz; $I$ peak 2.61 Jy bm$^{-1}$; lowest contour 2.09 mJy bm$^{-1}$; $p$ vectors clipped at 2.5 mJy bm$^{-1}$, (b) 7.9 GHz; $I$ peak 4.19 Jy bm$^{-1}$; lowest contour 3.77 mJy bm$^{-1}$; $p$ vectors clipped at 2.5 mJy bm$^{-1}$, (c) 12.9 GHz; $I$ peak 4.54 Jy bm$^{-1}$; lowest contour 4.08 mJy bm$^{-1}$; $p$ vectors clipped at 3.0 mJy bm$^{-1}$, (d) 15.4 GHz; $I$ peak 3.97 Jy bm$^{-1}$; lowest contour 5.56 mJy bm$^{-1}$; $p$ vectors clipped at 3.0 mJy bm$^{-1}$, (e) 22 GHz; $I$ peak 3.87 Jy bm$^{-1}$; bottom contour 9.67 mJy bm$^{-1}$; $p$ vectors clipped at 4.5 mJy bm$^{-1}$, (f) 43 GHz; $I$ peak 3.86 Jy bm$^{-1}$; lowest contour 38.62 mJy bm$^{-1}$; $p$ vectors clipped at 5.0 mJy bm$^{-1}$. The $I$ contours increase by factors of two in all maps. Length of $p$ vectors represent relative polarized intensity.}
 \label{1749_I}
\end{figure*}

\begin{figure}
\includegraphics[width=84mm]{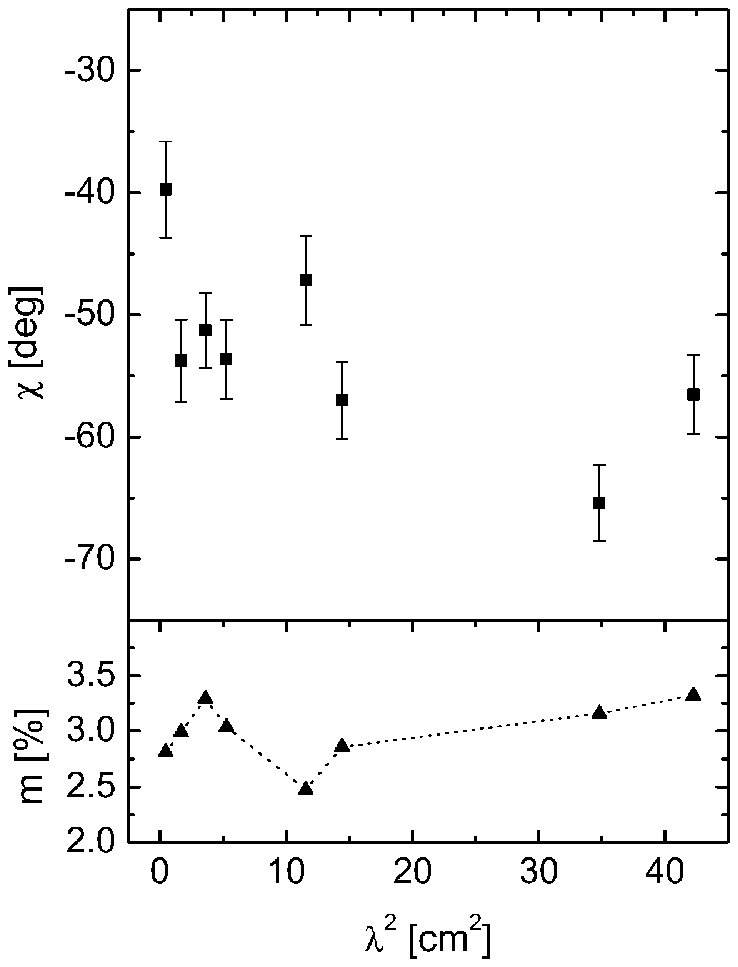}
 \caption{Plot of $\chi_{obs}$ vs. $\lambda^2$ for core of 1749+096. There is no evidence for a linear dependence indicating external Faraday rotation despite several attempts involving $n\pi$ rotations and $\pm90\degr$ rotations where significant changes in the spectral index were observed. The plot of the degree of polarization ($m$) underneath rules out the effect of internal Faraday rotation since $m$ remains approximately constant across the whole frequency range.}
 \label{1749angles}
\end{figure}

\subsubsection*{1749+096}
This BL Lac object has a redshift of 0.322 and is a point source on kpc scales \citep{rector2001}. On pc scales, it has a jet that extends in a north-easterly direction (Fig.~\ref{1749_I}), although the VSOP polarization observations of \citet{gabuzdavsop2003} show a new jet component emerging in a structural position angle of $\sim-15\degr$. This source has a relatively large integrated RM of $95\pm6$ rad m$^{-2}$ \citep{pushkarev2001}. The core region shows clear evidence for a spectral turnover, with the spectral index decreasing from $\alpha\sim+0.8$ between 4.6--8.9 GHz to $\alpha\sim-0.1$ between 12.9 and 43 GHz (Table \ref{spix}). Figure \ref{1749_I} displays the total intensity and polarization structure of 1749+096.

Reduced $\chi^2$ fits of the observed core polarization angles vs. $\lambda^2$ produced no acceptable fits across our entire frequency range for this source despite several attempts involving $\pm n\pi$ rotations and $\pm90\degr$ rotations where significant changes in the spectral index were observed (the latter due to possible optically thick--thin transitions); see Fig.~\ref{1749angles} for a plot of $\chi_{obs}$ vs. $\lambda^2$. No clear evidence of a $90\degr$ jump between the polarization angles at adjacent frequencies is observed, despite the evidence for a spectral turnover from the measured core spectral indices. This is surprising since previous observations have yielded good RM fits for this source \citep{zt2004, optical2006, mahmudalaska}. We may have observed the source during a flare since the peak flux for our observations is much larger than what has previously been observed. For example, our peak flux density at 7.9 GHz was 4.2 Jy/beam (epoch 2006), with previous observations that did detect Faraday rotation (epoch 2004) having a peak flux density of 3.0 Jy/beam at 7.9 GHz (Mahmud, private communication); also, previous 43-GHz observations \citep{optical2006} have a peak flux density of 2.4 Jy/beam whereas the peak flux density in our 43-GHz map is 3.9 Jy/beam. \citet*{sokolov2008} detected a correlation between the polarized flux in the 43-GHz core and the 15--43 GHz core RM in multi-epoch observations of BL Lac indicating significant interaction between the emitting plasma and the Faraday rotating material. Hence, if the rotating material is mainly in a boundary layer and if it was affected by a flare, this may explain the lack of a clear $\lambda^2$ fit for 1749+096 at this epoch.

The core degree of polarization remains relatively constant across the whole frequency range, with $m\sim3\%$ (Fig.~\ref{1749angles}, bottom) consistent with the birth of a new polarized component that dominates the emission. The absence of depolarization also rules out the effect of internal Faraday rotation in the core being responsible for the lack of a $\lambda^2$ dependence with polarization angle.
There was no appreciable jet polarization detected in this source. The core polarization orientation is approximately transverse to the visible VLBI jet at all frequencies (Fig.~\ref{1749_I}).

\subsubsection*{2007+777}
\begin{figure*}
\includegraphics[width=168mm]{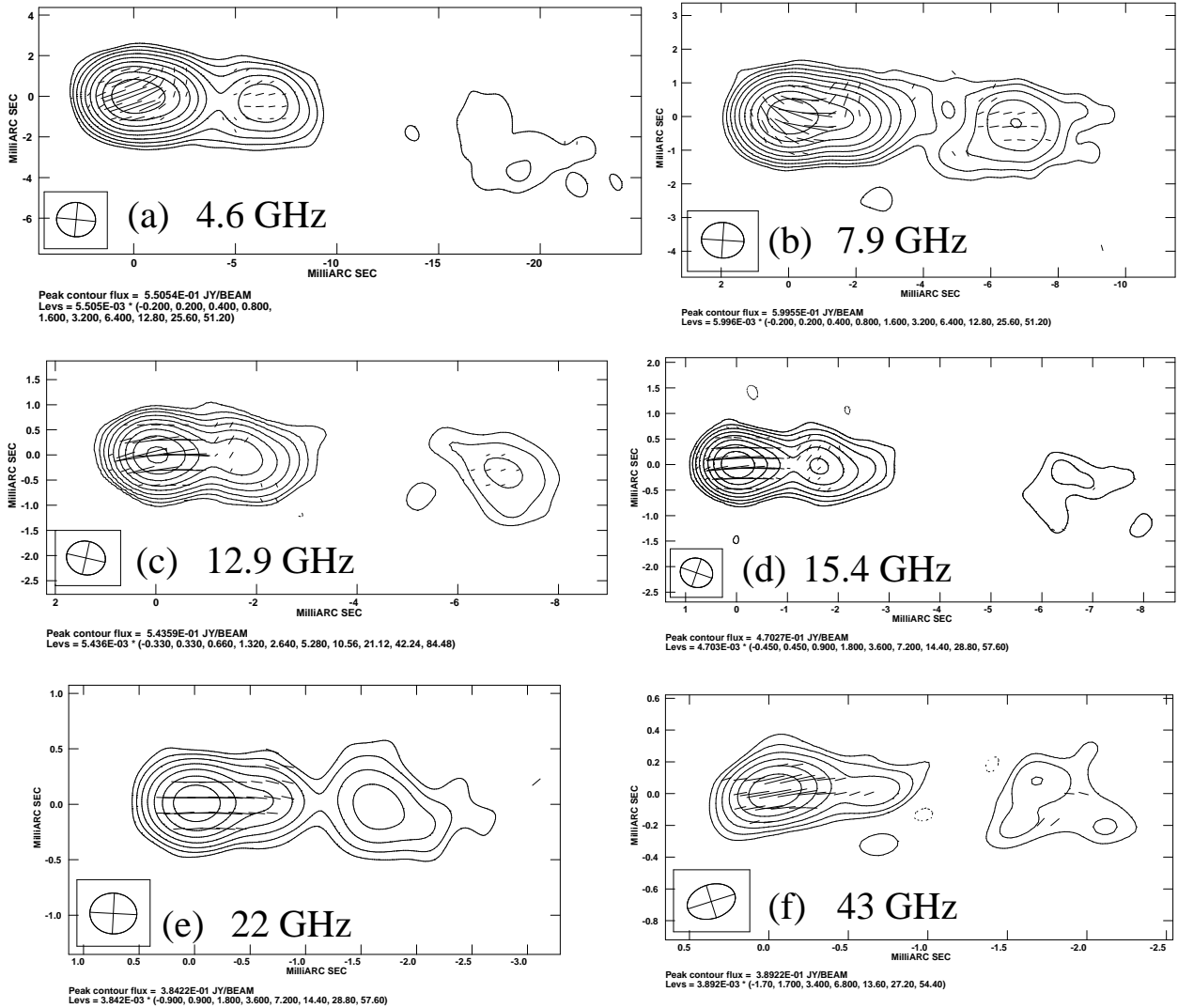}
 \caption{Total intensity ($I$) and polarization ($p$) maps for 2007+777: (a) 4.6 GHz; $I$ peak 0.55 Jy bm$^{-1}$; lowest contour 1.10 mJy bm$^{-1}$; $p$ vectors clipped at 1.0 mJy bm$^{-1}$, (b) 7.9 GHz; $I$ peak 0.60 Jy bm$^{-1}$; lowest contour 1.20 mJy bm$^{-1}$; $p$ vectors clipped at 1.0 mJy bm$^{-1}$, (c) 12.9 GHz; $I$ peak 0.54 Jy bm$^{-1}$; lowest contour 1.79 mJy bm$^{-1}$; $p$ vectors clipped at 1.5 mJy bm$^{-1}$, (d) 15.4 GHz; $I$ peak 0.47 Jy bm$^{-1}$; lowest contour 2.12 mJy bm$^{-1}$; $p$ vectors clipped at 1.5 mJy bm$^{-1}$, (e) 22 GHz; $I$ peak 0.38 Jy bm$^{-1}$; bottom contour 3.46 mJy bm$^{-1}$; $p$ vectors clipped at 5.5 mJy bm$^{-1}$, (f) 43 GHz; $I$ peak 0.39 Jy bm$^{-1}$; lowest contour 6.62 mJy bm$^{-1}$; $p$ vectors clipped at 7.5 mJy bm$^{-1}$. The $I$ contours increase by factors of two in all maps. Length of $p$ vectors represent relative polarized intensity.}
 \label{2007_I}
\end{figure*}

This source is classified as a BL Lac object with a redshift of 0.342. It has a two-sided kpc-scale jet \citep{antonucci1986b} orientated in the same direction as the one-sided VLBI pc-scale jet (Fig.~\ref{2007_I}). The integrated RM found by \citet{rusk} is $-20\pm3$ rad m$^{-2}$. Figure \ref{2007_I} displays the total intensity and polarization structure of 2007+777. The core is self-absorbed with $\alpha\sim+0.6$ from 4.6--8.9 GHz, while the jet is optically thin with $\alpha\sim-0.4$. From 12.9--43 GHz the core spectral index decreases to $\alpha\sim+0.2$ and $\alpha\sim-0.6$ in the jet (Table~\ref{spix}). The plot of $\chi_{obs}$ vs. $\lambda^2$ in Fig.~\ref{RMtransitions} is suggestive of an optically thick--thin transition between 8.9 and 12.9~GHz. This is supported by the large increase in the core degree of polarization between these frequencies (see Table~\ref{resultstable}), as well as the flattening of the core spectrum from $\alpha\sim0.6$ between 4.6 and 8.9 GHz to $\alpha\sim0.2$ from 12.9--43 GHz. Therefore, we should rotate the core EVPAs from 4.6 to 8.9 GHz by $-90\degr$ before comparing with the higher frequency EVPAs (see Fig.~\ref{RMtransitions}e).

\begin{figure*}
\includegraphics[width=162mm]{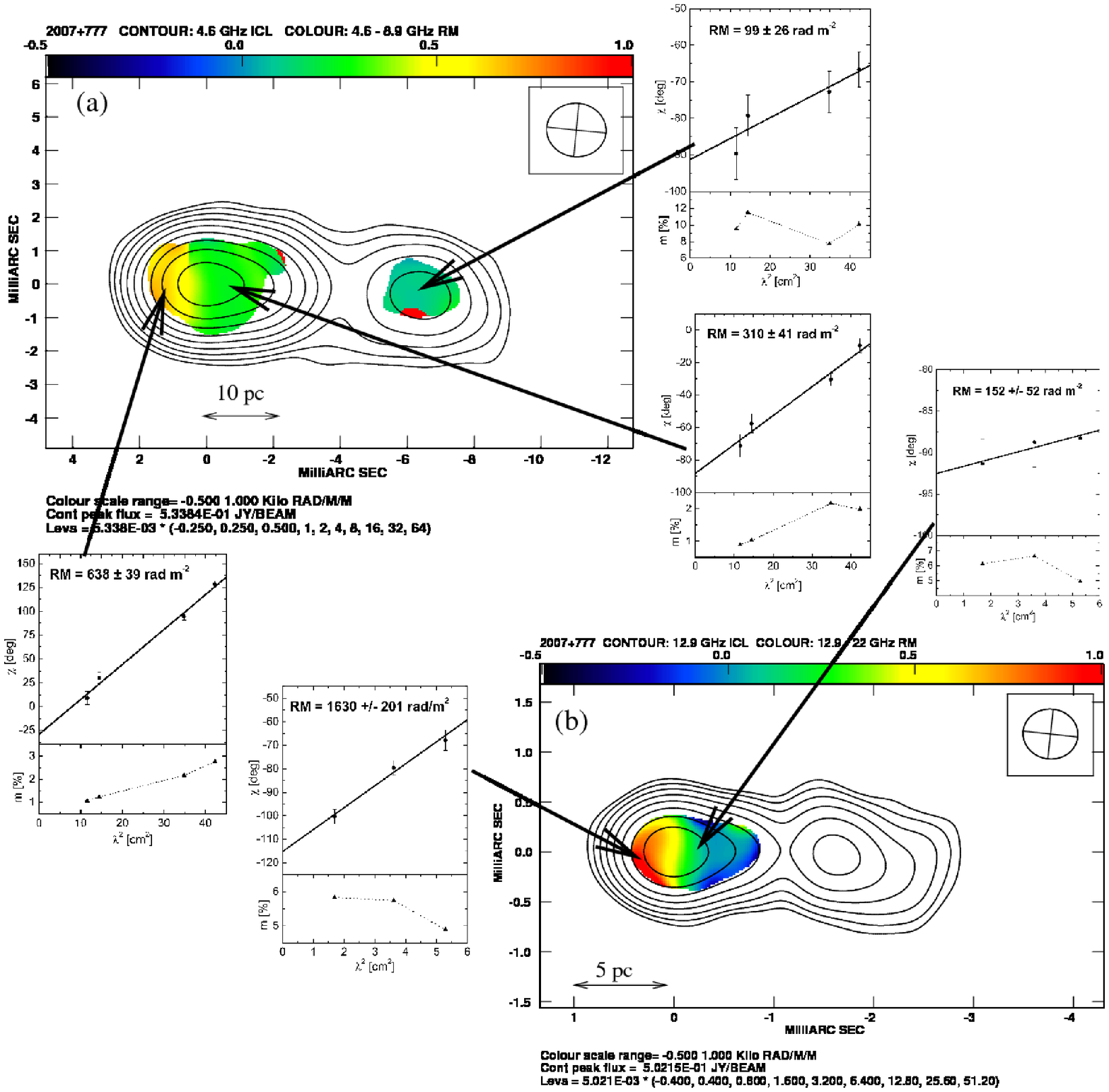}
 \caption{Source: 2007+777. (a) 4.6--8.9 GHz RM map (colour scale RM range: -0.5--1$\times10^3$ rad m$^{-2}$) with 4.6 GHz total intensity contours ($I$ peak 0.53 Jy bm$^{-1}$; lowest contour 1.33 mJy bm$^{-1}$). The solid arrow points to the core region where the $\chi$ vs. $\lambda^2$ fits shown in the inset were found; the degree of polarization is plotted underneath for each corresponding polarization angle. The dashed arrow points to the indicated slice taken across the RM map; the plotted black line shows the RM fit at every pixel across the slice with the grey area indicating the error in the fit for each point. (b) 12.9--22 GHz RM map (colour scale RM range: -0.5--1$\times10^3$ rad m$^{-2}$) with 12.9 GHz total intensity contours ($I$ peak 0.50 Jy bm$^{-1}$; lowest contour 2.00 mJy bm$^{-1}$). All $I$ contours increase by factors of two. Full resolution image in journal version.}
 \label{2007RM}
\end{figure*}

We constructed RM maps from 4.6 to 8.9 GHz and from 12.9 to 22 GHz. The 43-GHz core EVPA does not lie on the $\lambda^2$ line from 12.9 to 22 GHz, indicating that a physically different region is being probed. From 4.6--8.9 GHz, the RM map has a sharp boundary between the core and the inner jet (Fig.~\ref{2007RM}a); this sharp boundary is also present in the spectral index map. The core has an RM of $638\pm39$ rad m$^{-2}$ while the inner jet RM is lower at $310\pm41$ rad m$^{-2}$. Out in the jet the RM is lower again at $99\pm26$ rad m$^{-2}$.
From the 12.9--22 GHz frequency interval, the core RM increases to $1630\pm201$ rad m$^{-2}$ giving an estimate of $a\sim0.9$. The inner jet has a RM of $\sim0$ rad m$^{-2}$ within the errors (Fig.~\ref{2007RM}b).

The core degree of polarization decreases rapidly from $\sim6.5\%$ at 43 GHz to $\sim0.9\%$ at 8.9 GHz before rising slowly to $\sim2\%$ at 4.6 GHz. Where jet polarization is detected at lower frequencies, the degree of polarization remains relatively constant ($m\sim8$--$10\%$).

\begin{figure*}
\includegraphics[width=170mm]{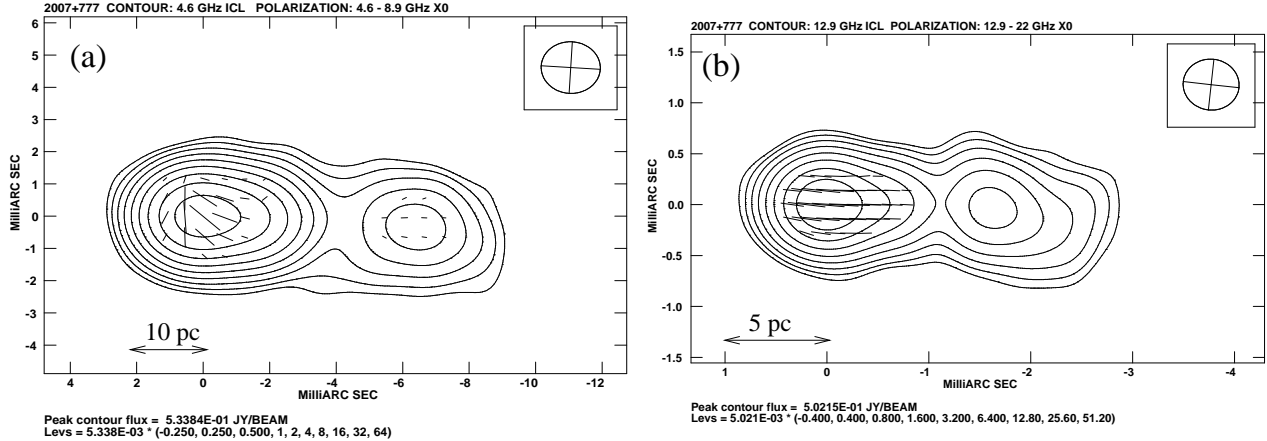}
 \caption{Total intensity ($I$) contours with intrinsic polarization orientation ($\chi_0$) for 2007+777 predicted from (a) 4.6--8.9 GHz RM map and (b) 12.9--22 GHz RM map. Same $I$ contour levels as in Fig.~\ref{2007RM}}
 \label{2007X0}
\end{figure*}

The Faraday corrected jet polarization is aligned with the jet direction in the 4.6--8.9 GHz $\chi_0$ map, while the core $\chi_0$ does not reflect the true $\chi_0$ distribution in this region because the polarized emission is dominated by optically thick regions at these frequencies (Fig.~\ref{2007X0}a). The inner jet and core $\chi_0$ values implied from the 12.9--22 GHz RM map are aligned with the jet direction (Fig.~\ref{2007X0}b). The 43-GHz core EVPA does not follow the $\lambda^2$ dependence from 12.9--22 GHz, possibly indicating that a physically separate RM region is being probed at this frequency.

\subsubsection*{2200+420}
This widely studied BL Lac object \citep[see][and references therein]{Bach2006} is the nearest source in our sample, with a redshift of 0.0686. It has been detected in $\gamma$-rays by EGRET \citep{mattox2001} and it has a kpc-scale core-halo structure \citep{antonucci1986a}.

\begin{figure*}
\includegraphics[width=168mm]{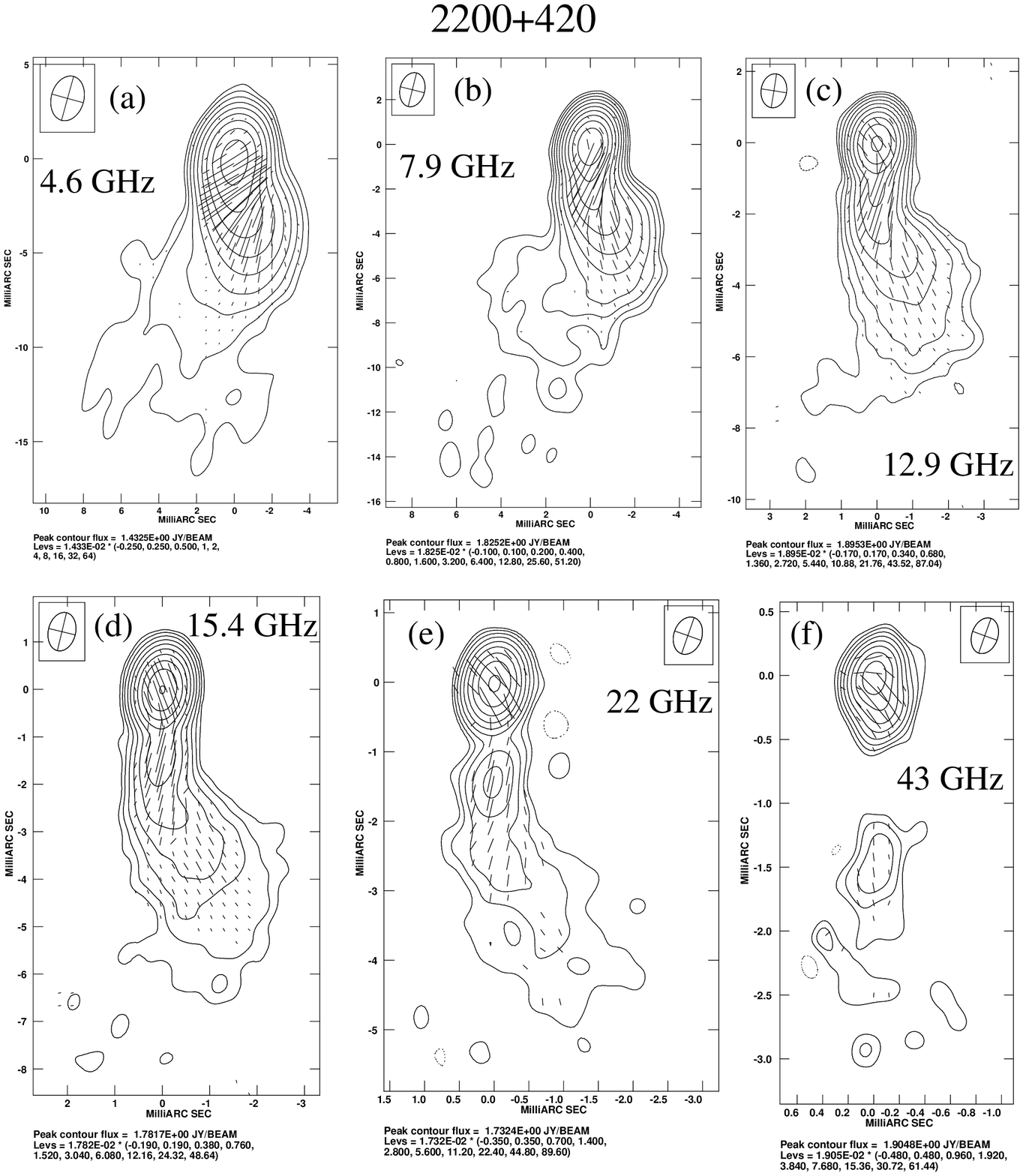}
 \caption{Total intensity ($I$) and polarization ($p$) maps for 2200+420: (a) 4.6 GHz; $I$ peak 1.43 Jy bm$^{-1}$; lowest contour 3.58 mJy bm$^{-1}$; $p$ vectors clipped at 1.5 mJy bm$^{-1}$, (b) 7.9 GHz; $I$ peak 1.83 Jy bm$^{-1}$; lowest contour 1.83 mJy bm$^{-1}$; $p$ vectors clipped at 1.5 mJy bm$^{-1}$, (c) 12.9 GHz; $I$ peak 1.90 Jy bm$^{-1}$; lowest contour 3.22 mJy bm$^{-1}$; $p$ vectors clipped at 1.8 mJy bm$^{-1}$, (d) 15.4 GHz; $I$ peak 1.78 Jy bm$^{-1}$; lowest contour 3.39 mJy bm$^{-1}$; $p$ vectors clipped at 2.0 mJy bm$^{-1}$, (e) 22 GHz; $I$ peak 1.73 Jy bm$^{-1}$; bottom contour 6.06 mJy bm$^{-1}$; $p$ vectors clipped at 4.0 mJy bm$^{-1}$, (f) 43 GHz; $I$ peak 1.90 Jy bm$^{-1}$; lowest contour 9.14 mJy bm$^{-1}$; $p$ vectors clipped at 5.5 mJy bm$^{-1}$. The $I$ contours increase by factors of two in all maps. Length of $p$ vectors represent relative polarized intensity.}
 \label{2200_I}
\end{figure*}

It has a large integrated RM of $-205\pm6$ rad m$^{-2}$ \citep{rudnickjones1983}. The core region is self-absorbed, with $\alpha\sim+0.7$ from 4.6--12.9 GHz and $\alpha\sim+0.5$ from 15.4--43 GHz. The extended jet emission is optically thin, with $\alpha\sim-0.5$ (Table \ref{spix}). Figure \ref{2200_I} displays the total intensity and polarization structure of 2200+420.

\begin{figure*}
\includegraphics[width=162mm]{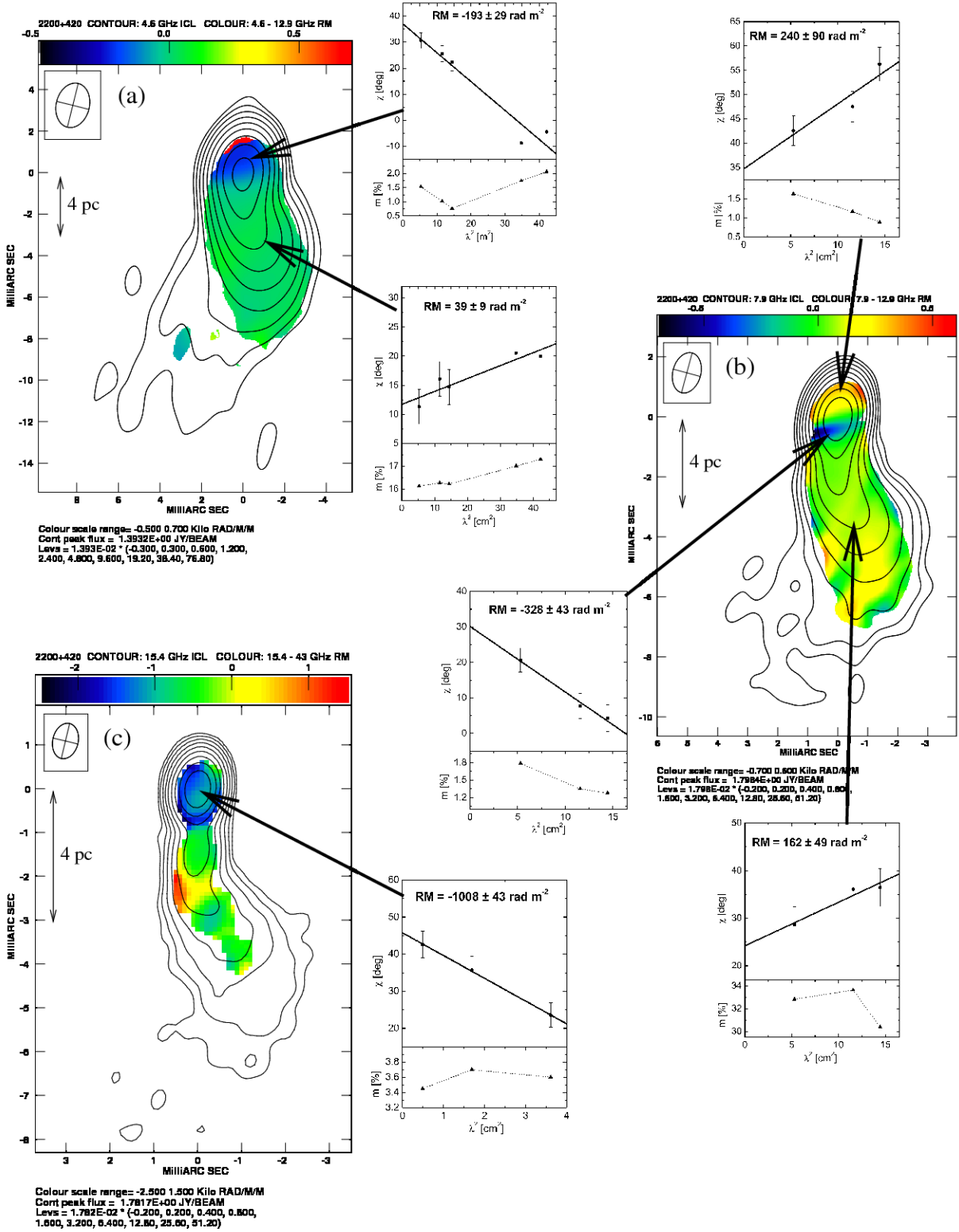}
 \caption{Source: 2200+420. (a) 4.6--12.9 GHz RM map (colour scale RM range: -0.5--0.7$\times10^3$ rad m$^{-2}$) with 4.6 GHz total intensity contours ($I$ peak 1.39 Jy bm$^{-1}$; lowest contour 4.18 mJy bm$^{-1}$). The solid arrow points to the core region where the $\chi$ vs. $\lambda^2$ fits shown in the inset were found; the degree of polarization is plotted underneath for each corresponding polarization angle. The dashed arrow points to the indicated slice taken across the RM map; the plotted black line shows the RM fit at every pixel across the slice with the grey area indicating the error in the fit for each point. (b) 7.9--12.9 GHz RM map (colour scale RM range: -0.7--0.6$\times10^3$ rad m$^{-2}$) with 7.9 GHz total intensity contours ($I$ peak 1.80 Jy bm$^{-1}$; lowest contour 3.60 mJy bm$^{-1}$). (c) 15.4--43 GHz RM map (colour scale RM range: -2.5--1.5$\times10^3$ rad m$^{-2}$) with 15.4 GHz total intensity contours ($I$ peak 1.78 Jy bm$^{-1}$; lowest contour 3.56 mJy bm$^{-1}$). All $I$ contours increase by factors of two. Full resolution image in journal version.}
 \label{2200RM}
\end{figure*}

Three separate RM maps were constructed from 4.6--12.9, 7.9--12.9 and 15.4--43 GHz. There was a clear transition in the $\lambda^2$ dependence of the core EVPA between 12.9 and 15.4 GHz, indicating a physically separate region of the jet was being probed (e.g., Fig.~\ref{RMtransitions}, bottom right). Interestingly, this transition also appears in the RM maps of BL Lac by \citet[][Fig. 25]{zt2003} but without the 22 and 43-GHz data it could not be identified.
After removing the integrated RM, we get a value of $39\pm9$ rad m$^{-2}$ out in the jet from the low resolution (4.6--12.9 GHz) RM map. This is consistent with a low electron density, as expected for the extended emission.
The core RM of $-193\pm29$ rad m$^{-2}$ is similar to previous values \citep{zt2003} with the integrated RM removed (Fig.~\ref{2200RM}a). The 4.6 and 5.1-GHz $\chi$ values do not fit exactly on the $\lambda^2$ line, possibly indicating a change in the sign of the RM at frequencies lower that 4.6 GHz. Upstream of the core, the RM map shows a small region of very high RM.
We were unable to prevent the {\sevensize AIPS} task {\sevensize RM} from unnecessarily rotating the 4.6 and 5.1 GHz EVPAs by $180\degr$ in this region, hence producing a spurious, large RM that is not consistent with the higher frequency observations.

Probing further down the jet in the 7.9--12.9 GHz frequency range, the sign of the core RM changes to $+240\pm90$ rad m$^{-2}$ (Fig.~\ref{2200RM}b). Because we only have three $\chi$ values in this range, the error is large, but the increase in the magnitude of the core RM is significant, and is consistent with the general trend of the RM magnitude increasing at higher frequencies.
Downstream of the core, a band of RM of $-328\pm43$ rad m$^{-2}$ is detected. These bands of RM with sign changes are a common feature of pc-scale RM maps \citep[e.g.,][]{zt2001} and are possibly associated with bends in the inner jet leading to changes in the dominant LoS component of the magnetic field (see Section 4.5).
The jet RM of $162\pm49$ rad m$^{-2}$ again supports a lower electron density in the jet compared to the core region, but is higher than the jet RM from the lower resolution 4.6--12.9 GHz data.

The highest frequency range (15.4--43 GHz) core RM of $-1008\pm43$ rad m$^{-2}$ changes sign again and increases in magnitude (Fig.~\ref{2200RM}c), giving a value of $a=1.40\pm0.18$. The RM fits in the jet are rather poor due to the weak polarized jet emission at 22 and 43 GHz. However, there is evidence for a significant increase in magnitude of the RM where the jet bends.

\begin{figure*}
\includegraphics[width=180mm]{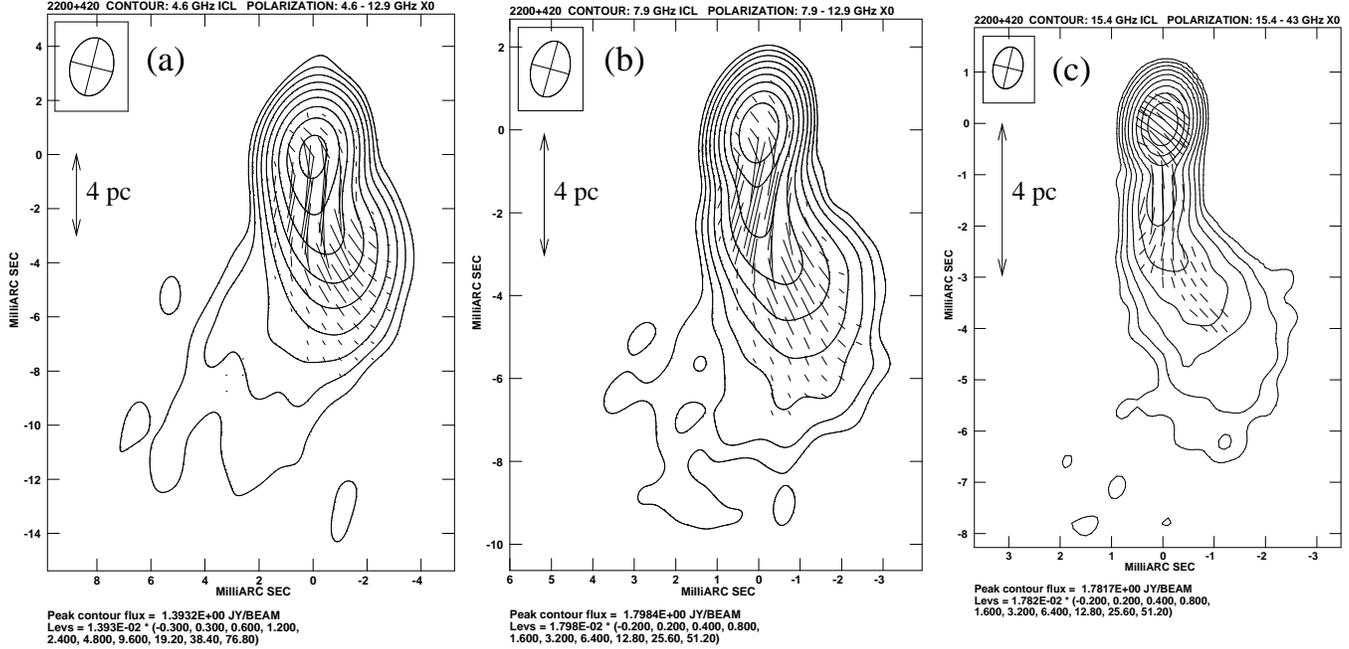}
 \caption{Total intensity ($I$) contours with intrinsic polarization orientation ($\chi_0$) for 2200+420 predicted from (a) 4.6--12.9 GHz RM map, (b) 7.9--12.9 GHz RM map and (c) 15.4--43 GHz RM map. Same $I$ contour levels as in Fig.~\ref{2200RM}}
 \label{2200X0}
\end{figure*}

The jet EVPAs are aligned with the jet direction after removing the effect of Faraday rotation, indicating a transverse magnetic field structure. There is significant interknot polarization and the intrinsic EVPAs remain well aligned with the jet direction even as it bends (Fig.~\ref{2200X0}a, b, c). The intrinsic core polarization orientation of $\sim40\degr$ appears to bear no relation to the observed inner jet direction in the 15--43 GHz $\chi_0$ map, but the higher resolution of our 43-GHz map shows a jet component in structural position angle $\sim210-220\degr$ from the core (Fig.~\ref{2200_I}f). Therefore, the EVPAs are also aligned with the jet direction in this region.

\begin{figure}
\includegraphics[width=70mm]{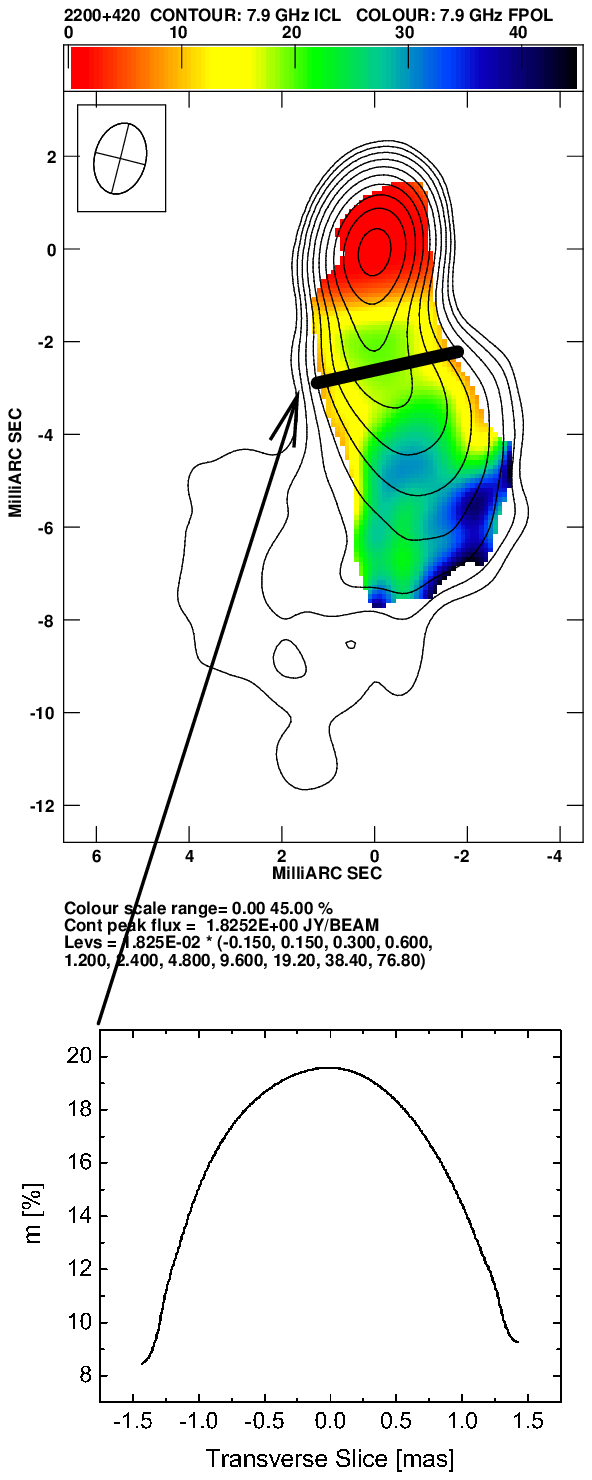}
 \caption{7.9 GHz total intensity and degree of polarization ($m$), ranging from 0 to 45 percent, for 2200+420 ($I$ peak 1.82 Jy bm$^{-1}$; bottom contour 2.74 mJy bm$^{-1}$). The sliced region shows the symmetric distribution of $m$ across the jet where it has a high $m$ along the spine of the jet and lower $m$ at the edges.}
 \label{2200mslice}
\end{figure}

The core degree of polarization decreases from $\sim2.1\%$ at 4.6 GHz to a minimum of $\sim1.3\%$ at 7.9 GHz before increasing to $\sim3.7\%$ at 22 GHz, with a slight decrease to $\sim3.5\%$ at 43 GHz.
The drop in core polarized flux at 43 GHz can be attributed to the newly resolved polarized component upstream of the core (Fig.~\ref{2200_I}f).
The degree of polarization in the jet remains relatively constant across the individual frequency ranges with $m\sim15$--$20\%$ at lower frequencies and increasing to values of $m\sim25$--$30\%$ at higher frequencies.
The degree of polarization maps show an unusual structure with the edges of the inner jet having lower $m$ ($\sim15\%$) than the central region with $m\sim20\%$ (Fig.~\ref{2200mslice}). Note that this is opposite to the trend expected for a helical jet magnetic field, which should yield increased $m$ at the jet edges.
This unexpected behaviour may be due to turbulence at the edges of the jet that disorders the magnetic field.

There is a substantial increase in $m$ where the outer jet bends. Even though the jet direction closer to the core has been observed to vary quite substantially \citep{stirling2003, muteldenn2004}, the outer jet has always been observed to bend to the south-east (e.g., Fig.~\ref{2200mslice}) and the higher $m$ at this bend possibly indicates that it is caused by a constant interaction with the surrounding medium.

\section{Discussion}

\subsection{Core Degree of Polarization}

The core degree of polarization has a V-shaped distribution for four (0954+658, 1156+295, 2007+777 and 2200+420) of the six sources in this sample, where it initially decreases rapidly from its value at the highest frequency and then, after reaching a minimum, increases again toward the lowest frequency (Fig.~\ref{depolfig}). We investigate whether Faraday depolarization across the observing beam could cause the observed decrease in $m$ using the model described in \cite{zt2004} and \citet{jorstad2007}, where $m(\%) = m_{\rm{RM=0}}|\rm{sinc}(RM\lambda^2)|$. We assume that the core degree of polarization at 43 GHz is $\sim90\%$ of $m_{\rm{RM=0}}$ as an initial estimate. For 0954+658 and 1156+295, the model has relatively good success in predicting the first null value in $m$. However, the degree of polarization at 22 GHz falls off faster than expected in both cases (see Figs.~{\ref{depolfig}a, b). For 2007+777 and 2200+420, the model describes the initial fall-off much better but underestimates $m$ at 7.9 and 8.9 GHz, although these values may be affected by blending with inner jet components (see Figs.~{\ref{depolfig}c, d). Therefore, we conclude that, while Faraday beam depolarization may have a significant effect, it cannot completely explain the initial fall-off in $m$.
The increase in the degree of polarization at low frequencies is most likely due to higher polarized inner jet regions becoming blended with the core, as described in \citet{gabuzdacawthorne1996} and \citet{gabuzda1999}.

It's also worth noting that this distribution is similar to the fractional linear polarization spectrum derived in \citet{jonesodell1977}, with a turnover frequency ($\nu_m$) for the core component dominating the observed polarization of $\sim$10--15 GHz. In their model the degree of polarization is high at frequencies much greater than $\nu_m$, then decreases rapidly to a minimum just below $\nu_m$ before increasing again toward low frequencies, similar to the sort of distribution we observe. To properly analyze this effect we need to find the spectral turnovers of individual model fitted components in the core; this will be investigated in a future paper.
With regard to the two sources which did not show this V-shaped distribution, the core $m$ for 1749+096 may be affected by the birth of a new strongly polarized component and for 1418+546 polarization was not detected at 22 or 43 GHz.

\subsection{Core Rotation Measure}

In all cases, the magnitude of the RM increases at higher frequencies, consistent with an increase in the electron density and/or magnetic field strength closer to the central engine. This is quantified using the relation $|\rm{RM_{core,\nu}}$$|$$\propto\nu^a$; see Table \ref{resultstable} for values of $a$.
Table \ref{resultstable} shows that two of our sources (1156+295 and 2007+777) are consistent with the value of $a\sim2$ expected for a stratified jet with a Faraday rotating boundary layer surrounding a conically expanding jet, similar to the results of \citet{jorstad2007}. However, 0954+658 and 1418+546 have significantly higher values of $a=3.3$ and 3.8, indicating much faster electron density fall-offs in the Faraday rotating medium with distance from the central engine. In the case of 0954+658, the fit is poor due to a large jump in the RM value at high frequencies (using only the two lower frequency core RMs we get an estimate of $a\sim1.8$), indicating that we are observing a region with much higher electron density. It is tempting to attribute the jump in core RM from 15--43 GHz to the affect of the broad-line region, however, \citet{blr2006} derive an upper limit on the size of the broad-line region for 0954+658 of $\sim0.4$ pc which is much smaller than the region probed by our 15-43 GHz observations of $\sim2$ pc. Another possibility is that the steep increase in the core RM is related to steep pressure gradients in the region confining the jet. \citet{lobanov1998} showed that the jet particle distribution follows approximately the same power-law as the external pressure at distances much greater than where the jet becomes supersonic using the hydrodynamic jet model of \citet{georganopoulosmarscher1998}. Hence, the large values of $a$ may correspond to jets confined by thermal pressure and/or ram pressure from a spherical disk-wind outflow surrounding a conically expanding jet \citep[e.g.,][]{bogovalovtsinganos2005}, while lower values of $a$ (e.g., $a=1.4$ for 2200+420) could be related to highly collimated outflows possibly with a significant contribution to confinment from magnetic pressure \citep{komissarov2007}.

The intrinsic core RM appears to be anti-correlated with the core degree of polarization (Fig.~\ref{intrinsicRM}), with a dearth of high polarized cores with large intrinsic RMs (with the exception being 2007+777 in the high frequency range where we observe an optical depth transition for the core polarization). This is consistent with the findings of \citet{zt2003} from a much larger sample for both quasars and BL Lac objects. They attribute this to depolarization from a gradient in Faraday rotation across the beam, but as we have seen this alone cannot completely explain the observed frequency dependence of $m$.

\subsection{Intrinsic Polarization Orientation}

After correction for Faraday rotation, the jet polarization orientation is approximately aligned with the jet direction for 0954+658, 2007+777 and 2200+420, and approximately transverse to the jet direction for 1418+546, while 1156+295 displays a spine-sheath polarization structure. EVPAs that are aligned with the jet direction have often been interpreted as being associated with transverse shocks that enhance the magnetic field component in the plane of compression \citep*[e.g.,][]{hughesaller1989}, but this interpretation has difficulty explaining significant aligned inter-knot polarization and aligned EVPAs following jet bends (e.g., Fig.~\ref{2200X0}). \citet*{lyutikov2005} have shown that this bi-modal distribution of the jet EVPAs is consistent with the presence of a helical magnetic field geometry, where the aligned EVPAs correspond to the dominant toroidal component of a relatively tightly wound helical field and the transverse jet EVPAs correspond to the poloidal component of a relatively loosely wound helical magnetic field. Regions of longitudinal magnetic field near the jet edges (Fig.~\ref{1156_I}a) are often attributed to interaction between the jet and the surrounding medium \citep[e.g.,][]{attridge2005}, but in fact can also be explained as the projected longitudinal component of a helical magnetic field at the jet edges, when there is sufficient resolution across the jet.

The intrinsic core EVPAs are more complicated to analyse due to possible blending of polarized components in the innermost jet at low frequencies. They also vary with frequency-interval because the different core RMs can imply different zero-wavelength angles, due to the different regions being probed. 2200+420 is the only source whose intrinsic core EVPA maintains an approximately constant direction, with $\chi_{0}(4.6$--$12.9 \rm{GHz})=37\pm4\degr$ and $\chi_{0}(15.4$--$43 \rm{GHz})=46\pm1\degr$, which is aligned with the inner jet direction seen at 43 GHz implying a dominant transverse magnetic field all along the pc-scale jet. For all other sources, we use the highest frequency-interval core $\chi_0$ value or, if the RM fits are unreliable or the core region is affected by blending, we use the 43-GHz core polarization angle.
For 0954+658, $\chi_{43 \rm{GHz}}\sim30\degr$ which is approximately transverse to the inner jet direction of $\sim-50\degr$. It is unclear whether the magnetic field is parallel or perpendicular to the core EVPA since we are not sure if the polarized emission is coming predominately from an optically thick or thin emitting region within the observed VLBI ``core''.
In the case of 1156+295, the value of $\chi_{0}(12.9-43 \rm{GHz})=-17\pm4\degr$ is offset from the observed inner jet direction of $\sim0\degr$, but a jet direction of $\sim-17\degr$ at higher frequencies would not be inconsistent with the helical jet trajectory proposed by \citet{hong2004}.
The highest frequency RM correction for the core of 1418+456 gives $\chi_{0}(7.9-15.4 \rm{GHz})=109\pm3\degr$ compared to a inner jet direction of $\sim130\degr$. The approximate $90\degr$ change in the EVPA between the core and inner jet suggets that the observed core polarization at high frequencies may be from the optically thick region, whereas the observed core polarization at lower frequencies is dominated by polarized emission from the optically thin inner jet region. In this case, the magnetic field direction would be approximately longitudinal in both the core and jet. The core and inner jet spectral indices for 1418+546 of $+0.3$ and $-0.6$, respectively, support this hypothesis, as well as the degree of polarization at 15 GHz, which increases from $\sim2\%$ in the core to $\sim10\%$ in the inner jet. The 43-GHz core polarization angle for 1749+096 of $\sim-50\degr$ is perpendicular to the jet direction of $\sim40\degr$. It's unclear whether the dominant magnetic field structure is transverse or longitudinal in this source, since the EVPAs remain approximately transverse to the jet direction across the entire frequency range, while the spectral index indicates that the core is strongly self-absorbed at low frequencies ($\alpha\sim+0.8$) but turns over at high frequencies with $\alpha\sim-0.1$. Using the 43-GHz core polarization angle for 2007+777, we find the core EVPA of $\sim-80\degr$ approximately aligned with the jet direction of $-90\degr$. This indicates that this source is dominated by a transverse magnetic field structure in both its pc-scale core and jet.

\subsection{Transverse Jet Structure}
Analysis of the transverse RM structure of 0954+658, 1418+546 and 1156+295 reveals the presence of positive asymmetric gradients in RM across the jet, which are strong signatures for the presence of a helical magnetic field geometry in the Faraday rotating medium. RM gradients in 0954+658 and 1156+295 have previously been reported by \citet{mahmudalaska} and \citet{gabuzdacp2008} on similar scales, and 0954+658 also has an RM gradient on decaparsec scales \citep{hallahanparis}. For a perfectly side-on view of a helical magnetic field, one would expect a symmetric gradient, with the LoS magnetic field changing sign across the jet. However, as was shown in \citet{asada2002}, a positive asymmetric gradient (as detected in 0954+658, 1418+546 and 1156+295) can be understood if the viewing angle in the jet rest frame is less than the pitch angle of the helix. All the observations show that the RM gradients in these three sources maintain a constant direction across the jet.

An excellent diagnostic of the internal magnetic field structure are transverse profiles of the synchrotron emission from the jet. However, determining the ``ridge line'' along the jet from which to take orthogonal slices can prove rather difficult. Using model fitted total intensity components along the jet is not very useful because some jets have quite broad funnels through which components move following different paths. Since the transverse jet emission itself can be asymmetric, using the peak intensity with distance along the jet is also not always appropriate. One option is to estimate the jet edges and define the ridge line as the mid point. However, estimating the jet edges where the emission is weak is often difficult and error prone. For the transverse profiles presented in this paper, we defined the normal to jet direction ``by eye'' and then checked that the profiles did not change dramatically for small changes of the slice direction, so that our conclusions didn't depend on our choice of profile.

Asymmetric distributions of transverse profiles of total intensity and polarization appear to be a common feature in pc-scale jets (e.g., Fig.~\ref{slices}). Helical magnetic fields produce asymmetric total intensity and polarization transverse profiles unless the jet is viewed at exactly $90\degr$ in the jet rest frame \citep{laing1981}. Models of a magnetic field structure in a cylindrical jet containing a uniform helical field with a tangled component (to reduce the degree of polarization to observed levels) are able to reproduce the observed offsets between the peaks in the total and polarized intensities (Papageorgiou \& Cawthorne 2008, in prep.) providing strong support for a helical magnetic field in the synchrotron emitting plasma.

Furthermore, the total intensity is maximum where the component of the magnetic field is perpendicular to the LoS ($I\propto\rm{Sin}[\theta]$), while the RM is maximum where the component of the magnetic field is parallel to the LoS (RM $\propto$ Cos[$\theta$]). Hence, for a helical magnetic field viewed at an angle other than $90\degr$ in the jet rest frame, the magnitude of the RM should be greatest on the opposite side of the jet from the total intensity peak. In the case of 0954+658 (Fig.~\ref{slices}, a) and 1156+295 (Fig.~\ref{slices}, b), it is not clear as the offset from the centre of the jet is quite small, $\sim6\%$ and $\sim3\%$ of the their respective beam-sizes, but this is not the case for 1418+546 (Fig.~\ref{slices}, c) where the offset from the centre of the jet to the $I$ peak is $\sim20\%$ of the beam. Since we do not expect the thermal plasma causing the observed Faraday rotation to be mixed in with the synchrotron emitting material, the RM and $I$ slice may not correspond to exactly the same helical field structure. If the jet is electromagnetically dominated, then the field lines must close back in the accretion disk, and it's possible that the toroidal component of the field in the sheath surrounding the jet is orientated in the opposite direction, causing the magnitude of the RM to be largest on the same side of the jet as the peak offset of the synchrotron emission (Figs.~\ref{slices}, b and c). However, comprehensive modelling of these profiles is clearly required to reach any firm, quantitative conclusions.

\begin{figure*}
\includegraphics[width=168mm]{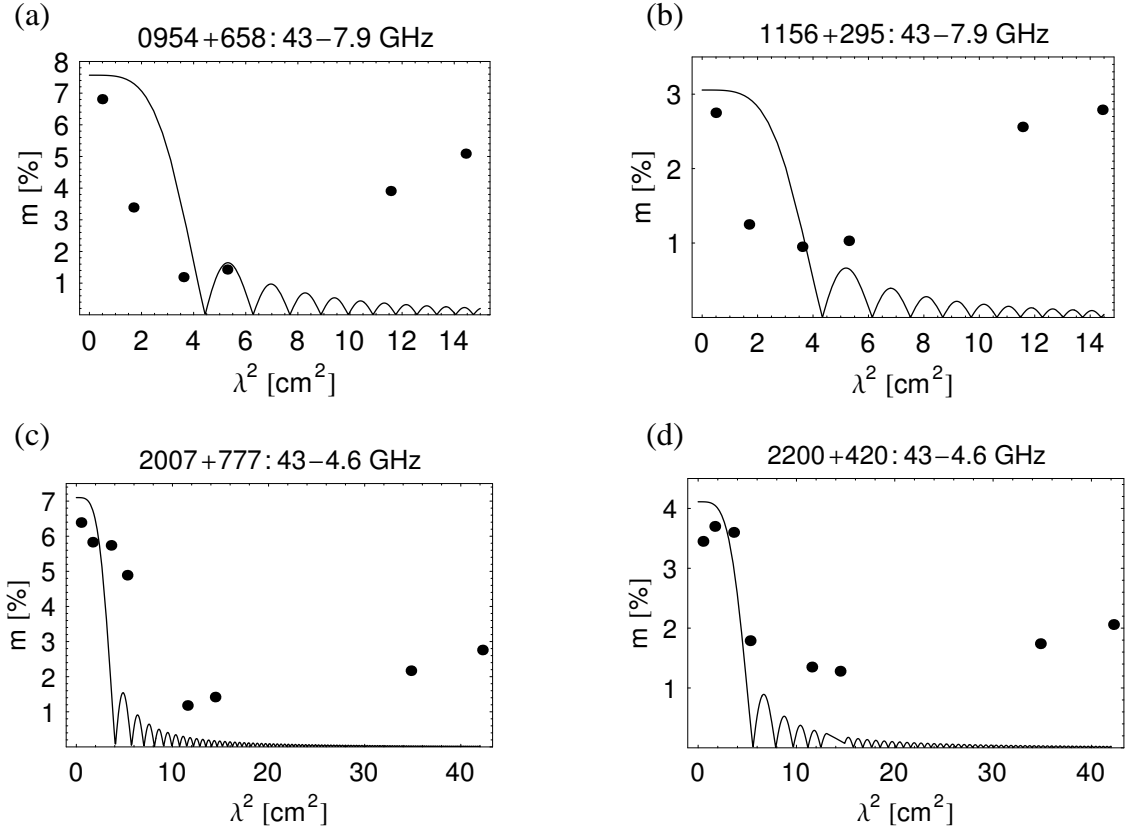}
 \caption{Core degree of polarization vs. $\lambda^2$ from 43--7.9 GHz for (a) 0954+658 and (b) 1156+295, and from 43--4.6 GHz for (c) 2007+777 and (d) 2200+420. Solid line shows predicted beam depolarization. See Section 4.1 for details.}
 \label{depolfig}
\end{figure*}

\subsection{Core RM Sign Changes}
RM sign changes in different frequency-intervals are observed in 0954+658, 1418+546 and 2200+420. In the case of 0954+658 and 2200+420, the dominant jet magnetic field is transverse throughout the jet (even as the jet bends, in the case of 2200+420) implying the presence of a global transverse magnetic field structure.
If we suppose that a helical magnetic field surrounds the jet, the observed RM sign reversals can be explained by slight bends of the relativistic pc-scale jet, for example, due to a collision with material in the parent galaxy or some instabilities inherent in the jet itself \citep[see][]{agudo2008}.

To show this, we generated simple model transverse RM profiles considering a helical magnetic field threading a sheath of Faraday rotating material surrounding the jet. On the left panels of Fig.~\ref{bends}, the top side of the jet, which corresponds to negative values on the transverse profile axis, has the toroidal component of the helical field orientated towards the observer. Relativistic aberration of the radiation emitted by the jet means that radiation emitted at $90\degr$ in the jet rest frame is observed at an angle ($\theta$) of $\sim1/\Gamma$ in the observers frame, where $\Gamma$ is the bulk Lorentz factor of the jet.
Hence, a side-on view ($\theta\sim1/\Gamma$) of a helical magnetic field (Fig.~\ref{bends}, top left) will have an RM that varies uniformly across the jet from positive to negative; considering an unresolved transverse jet structure, a zero net RM should be observed.
If the jet bends towards us, we have a ``head-on'' view of the helical magnetic field (i.e., $\theta < 1/\Gamma$), providing an asymmetric RM profile across the jet (Fig.~\ref{bends}, middle left). The dominant LoS magnetic field component comes from the top half of the jet and, for a beam-size comparable to or bigger than the diameter of the jet, the transverse RM structure is resolved out and a net positive RM will be observed in this case.
Conversely, for a ``tail-on'' view of a helical magnetic field (i.e., $\theta > 1/\Gamma$) (Fig.~\ref{bends}, bottom left), a negative RM will be observed.
(A longitudinal jet magnetic field with a change in the angle to the LoS could also cause the observed effects, but we have considered a helical field here, because the observed intrinsic EVPAs usually indicate a dominant {\em transverse} magnetic field and all three sources display RM gradients that provide direct evidence for helical fields associated with their jets.)

The RM sign changes could also be due to acceleration/deceleration of the jet since this will change $\Gamma$ and therefore, the viewing angle relative to $1/\Gamma$. Recent observations \citep[e.g.,][]{jorstad2005} have provided evidence for pc-scale acceleration in blazar jets, and \citet{vlahakiskonigl2004} show that an extended (i.e., pc-scale) acceleration region is a distinguishing characteristic of MHD driving of relativistic outflows compared to purely hydrodynamic models. Importantly, our explanation for the RM sign changes implies relativistic bulk motions of the Faraday rotating material in the sheath, at least in the innermost part of the jet. This would support any theoretical simulations indicating a mildly relativistic outflow from the accretion disk surrounding the radio emitting part of the jet.

Therefore, it's possible that either pc-scale bends or an accelerating/decelerating jet surrounded by a helical magnetic field could be responsible for the observed RM sign changes, because both effects cause the viewing angle to change relative to the angle $1/\Gamma$ along the jet.
As noted in Section 2.1, VLBI resolution is usually not sufficient to completely resolve the true optically thick core, so on scales smaller than the observed VLBI core, ``core'' RMs with different signs could be derived from observations at different frequencies (i.e., probing different scales of the inner-jet).

\begin{figure}
\includegraphics[width=84mm]{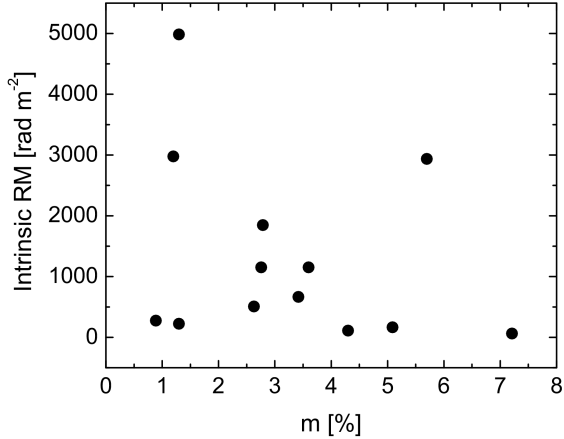}
 \caption{Plot of intrinsic core RM ($\rm{RM_{intrinsic}}=\rm{RM_{observed}}(1+z)^2$) for all sources vs. the corresponding core degree of polarization ($m$) listed in Table~\ref{resultstable}. There is clearly a lack of highly polarized cores with large core RMs.}
 \label{intrinsicRM}
\end{figure}

\begin{figure}
\includegraphics[width=84mm]{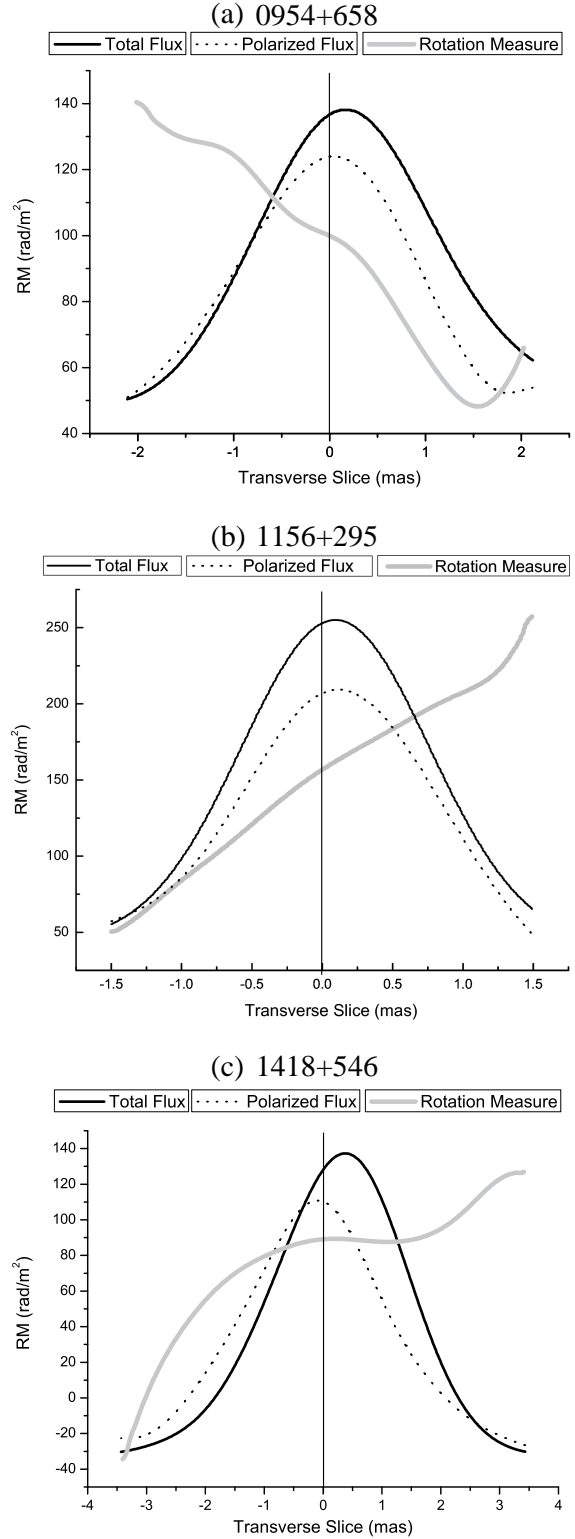}
 \caption{(a) Transverse profiles of the 4.6 GHz total intensity and polarized intensity of 0954+658 along with the 4.6--15.4 GHz RM slice across the region indicated in Fig.~\ref{0954RM}a. (b) Transverse profiles of the 4.6 GHz total intensity and polarized intensity of 1156+295 along with the 4.6--15.4 GHz RM slice across the region indicated in Fig.~\ref{1156RM}a. (c) Transverse profiles of the 4.6 GHz total intensity and polarized intensity of 1418+546 along with the 4.6--8.9 GHz RM slice across the region indicated in Fig.~\ref{1418RM}a.}
 \label{slices}
\end{figure}

\section{Conclusions}

We have analysed the polarization and RM distribution of six blazars from 5--43 GHz and a summary of our results is outlined below.

1. We find that the magnitude of the core RM increases systematically with increasing frequency and is well described in most cases by $|\rm{RM_{core}}$$|$$\propto\nu^{a}$. The values of $a$ vary from 0.9 to 3.8. Intermediate values of $a\sim2$ would be expected for a boundary layer of Faraday rotating material surrounding a conically expanding jet. Low values of $a$ may correspond to highly collimated outflows while large values of $a$ are possibly related to large external pressure gradients from a spherical disk-wind type geometry.

2. The core degree of polarization for our observed sources decreases rapidly from its value at 43 GHz, to a minimum somewhere between 8--15 GHz before increasing again at lower frequencies. The initial decrease cannot be completely explained by Faraday beam depolarization, and we attribute the low frequency increase to blending of higher polarized inner jet components.

3. We detect gradients in the RM across the jets of 0954+658, 1156+295 and 1418+546, which provides evidence for the presence of helical magnetic fields in a sheath or boundary layer surrounding the jet.

4. After correcting the polarization vectors for the effect of Faraday rotation, we find the jet EVPAs to be approximately aligned with the jet direction for 0954+658, 2007+777 and 2200+420. 1156+295 displays a spine-sheath polarization structure, and in 1418+546 the intrinsic jet EVPAs are approximately transverse to the jet direction. The EVPAs of 2200+420 display a continuous structure that remains aligned with the jet direction even as it bends. This shows that the magnetic field structure in the synchrotron emitting plasma is dominated by an ordered transverse component in all of these sources except for 1418+546. A helical magnetic field geometry can neatly explain both the bi-model distribution of the jet EVPAs and the well ordered polarization structures.

5. For three of the five sources with detected Faraday rotation (0954+658, 1418+546 and 2200+420), we find that the core RM also changes sign with distance from the central engine. We provide an explanation for this by considering a boundary layer of Faraday rotating material threaded by a helical magnetic field, with bends in the relativistic jet or acceleration/deceleration of the jet leading to changes in the viewing angle relative to the angle $1/\Gamma$ (corresponding to a viewing angle of $90\degr$ in the jet rest frame). This in turn gives rise to changes in the sign of the dominant LoS component of the magnetic field, and thereby to changes in the sign of the RM.

%

\begin{figure*}
\includegraphics[width=162mm]{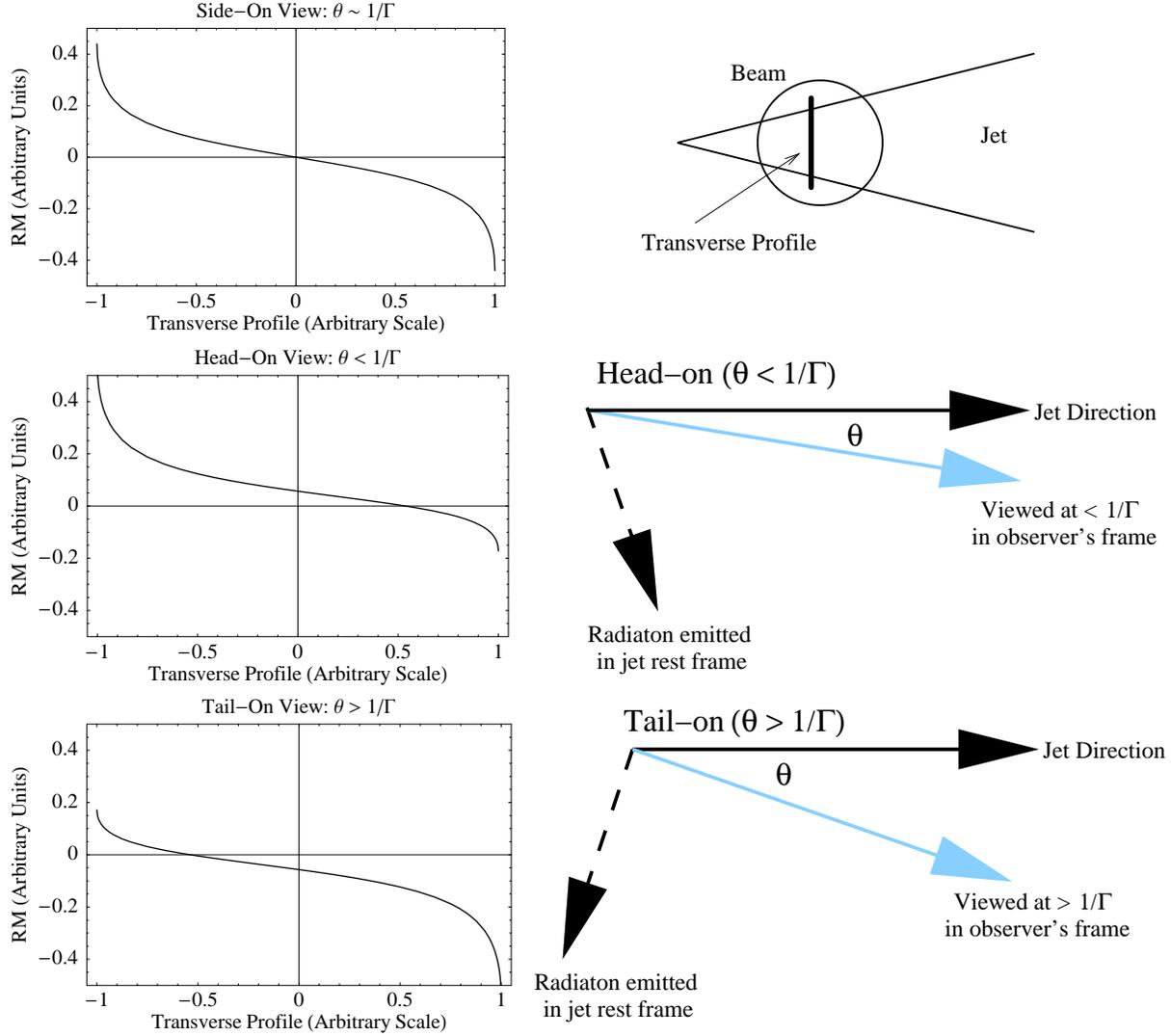}
 \caption{Top right: Cartoon of a conical jet with the heavy line indicating a transverse profile of the jet. The circle is an example of the finite resolution of our VLBI observations. Middle right: Due to the relativistic motion of the jet towards us, radiation emitted at $<90\degr$ in the jet rest frame is observed at $<1/\Gamma$ where $\Gamma$ is the bulk Lorentz factor of the jet. Bottom right: Radiation emitted at $>90\degr$ in the jet rest frame is observed at $>1/\Gamma$. Top left: Transverse RM profile of a jet surrounded by a sheath of Faraday rotating material and viewed at $90\degr$ in the jet rest frame. Due to the finite resolution of our observations a zero net RM would be observed. Middle left: Transverse RM profile for a jet viewed ``head-on'' ($< 90\degr$ in the jet rest frame). If we cannot resolve the transverse jet structure a net positive RM will be observed. Bottom left: Transverse RM profile for a jet viewed ``tail-on'' ($> 90\degr$ in the jet rest frame). In this case, a net negative RM will be observed if the transverse structure is resolved out.}
 \label{bends}
\end{figure*}

\bibliographystyle{mn2e}
\bibliography{osullivangabuzda}

\bsp
\end{document}